\documentclass[twocolumn,superscriptaddress,showpacs,aps,prb,floatfix]{revtex4-2}
\usepackage{amsmath,amsfonts,amssymb,epsfig,dcolumn,bm,graphics,latexsym,color,graphicx,multirow,ulem}
\usepackage[utf8]{inputenc}
\usepackage[T1]{fontenc}

\def\\ri{\mathrm{i}}

\let\oldAA\AA

\def\ri{\mathrm{i}}
\renewcommand{\AA}{\text{\normalfont\oldAA}}
\begin{document}
\preprint{AIP/123-QED}
\title{
Strain-tuning of spin anisotropy in single-layer phosphorene: insights from Elliott-Yafet and Dyakonov-Perel spin relaxation rates
}
\author{Paulina Jureczko}
\email{paulina.jureczko@us.edu.pl}
\affiliation{Institute of Physics, University of Silesia in Katowice, 41-500 Chorzów, Poland }
\author{Marko Milivojevi\' c}
\affiliation{Institute of Informatics, Slovak Academy of Sciences, 84507 Bratislava, Slovakia}
\affiliation {Faculty of Physics, University of Belgrade, 11001 Belgrade, Serbia}
\author{Marcin Kurpas}
\email{marcin.kurpas@us.edu.pl}
\affiliation{Institute of Physics, University of Silesia in Katowice, 41-500 Chorzów, Poland }
\begin{abstract}
Materials and systems that exhibit persistent spin texture provide  a platform for creating robust spin states that can be used in quantum computing, memory storage, and other advanced technological applications. 
In this paper we show that persistent spin-texture in single-layer phosphorene electrons close to the $\Gamma$ point, subjected to the finite perpendicular electric field, can be achieved by appropriately tuning the extrinsic spin-orbit coupling strength using the tensile strain of about $1.2\%$ in the zigzag direction. 
This is confirmed by detailed numerical investigations of the effects of strain on the intrinsic and extrinsic spin-orbit coupling, and by the effective spin-orbit Hamiltonian of phosphorene electrons and holes around the $\Gamma$ point, assuming the presence of the perpendicular electric field. Furthermore, the calculated spin relaxation rates due to the Dyakonov-Perel mechanism indicate a giant anisotropy of the in-plane spin, up to $10^5$, which is directly related to the discovered persistent spin texture of phosphorene electrons close to the $\Gamma$ point. 
We also show, that strain can reverse the anisotropy of spin mixing parameter $b^2$ connected to the Elliott-Yafet spin relaxation mechanism which dominates spin relaxation in phosphorene. We find the conditions under which Elliott-Yafet spin lifetime anisotropy can be largely enhanced due to synergy of spin mixing and g-factor anisotropy.
Our results suggest that spin texture in phosphorene can be modulated by strain, enabling its potential usage in the field of spintronics.  
\end{abstract}
\maketitle
\section{Introduction}
Spintronics~\cite{W+01,ZFS04,FME+04,HKG+14,J12} is a subfield of electronics focused on manipulation of electrons' spin  in addition to their charge. Exchange coupling and spin-orbit coupling (SOC) represent two fundamental interactions in spintronics that play crucial roles in controlling the spins of electrons. These interactions provide mechanisms to manipulate and transport spin information, which is essential for developing advanced spintronic devices.

Phosphorene has great potential for electronics applications due to its high carrier mobility, tunable band gap, and anisotropic properties~\cite{LNZ+14,LNH+14,LWL+14,ZYX+14,CVP+14,QKH+14}.
Weak spin-orbit coupling of phosphorus atoms promotes nanosecond spin lifetimes in phosphorene \cite{ATK+17,CLT+24}, which, combined with the intrinsic semiconducting band gap, positions phosphorene as an excellent spintronic material. 
However, spin degenerate band structure of phosphorene, present due to the structure inversion symmetry limits its spectrum of applications, since individual spin states cannot be addressed individually. 
The spin degeneracy of bands can be removed by applying an external transverse electric field~\cite{KGF16,PFG+24} and/or by placing the phosphorene on the substrate with sizable SOC, for example WSe$_2$~\cite{MGK+23,MGK+24,MKR+24}.

Whereas both the electric-field and proximity-induced manipulation are effective in tuning the SOC strength in phosphorene electrons and holes~\cite{WP14,HFM+23}, it is not possible to control the spin anisotropy very effectively.

Anisotropy is intrinsic to phosphorene.  
Like its bulk counterpart, monolayer black phosphorus has an orthorhombic crystal structure belonging to the  $C_{mca}$ space group, being isomorphic with non-symmorphic $D_{2h}$ point group.  
The $sp^3$ hybridization of atomic orbitals leads to a characteristic puckering of the crystal structure, which splits off the planar lattice into two connected atomic zigzag chains at different heights Fig.\ref{fig:structure} a).  The bonding energy is dominated by in-plane bonds within the zigzag chains \cite{Li2014}, while the out-of-plane bonds connecting the zigzag chains along the armchair direction are energetically less stable. This imbalance in bonding energy and anisotropic crystal structure broadcast the anisotropy to almost all properties of phosphorene, starting from mechanical \cite{Wang2015}, thermal \cite{Luo2015},  electronic properties \cite{LNZ+14,QKH+14} up to optical activity \cite{Wang2015_exc} and spin properties \cite{KGF16,ATK+17,CLT+24}.

In this manuscript, using first-principles calculations and effective models, we explore the role of strain in shaping spin anisotropy of phosphorene electrons and holes. To trace the effects of strain on intrinsic spin-orbit coupling we calculate the spin mixing parameter $b^2$ for spin degenerate Bloch states.  We show that for both in-plane  and out-of-plane spin direction, the spin mixing parameter is weakly sensitive to uniaxial strain up to 4\%, giving almost constant out-of-plane to in-plane ratio $b^2_{\perp}/b^2_{\parallel}\sim 2$. When the applied strain approaches 5\,\%, spin mixing anisotropy for conduction electrons reverses due band anticrossing. In the doping range up do 5\,meV  we observe a strong anisotropy, up to $b^2_{\parallel}/b^2_{\perp}\sim 100$, which decreases to an approximately constant ratio $b^2_{\parallel}/b^2_{\perp}\sim 10$ for higher doping values. 

In the case of nonzero electric fields, effects of strain on the spin splitting and spin texture of phosphorene electrons and holes are analyzed using first principles calculations and an effective spin-orbit Hamiltonian.
We show that by applying a suitable ($\propto1.2\%$) compressive strain in the zigzag direction of the phosphorene monolayer, it is possible to generate conditions for the persistent spin texture in the conduction band with only $S_x$ spin component. This leads to a giant, $\sim 10^5$, anisotropy of extrinsic  spin-orbit fields $\Omega$ and leads to a strong damping of Dyakonov-Perel spin relaxation for $S_x$ spins.  In contrast to $b^2$, the anisotropy direction of $ \Omega$  does not change with doping.

Until now, strain has been recognized as a successful tool for tuning the electronic properties \cite{Rodin2014, Elahi2015} and, phonon spectra \cite{Ge_2015,Elahi2015}. We show that tuning spin properties of phosphorene by strain is possible, effectively transforming it into a material suitable for advanced technological applications in quantum computing and/or memory storage.


The paper is organized as follows. In Sec.~\ref{DFTbasics} we present the geometry of the strained phosphorene monolayer under the presence of the perpendicular electric field, and provide necessary numerical details about the performed first-principles calculations. In Sec.~\ref{b2parameter} the analyze the effects of strain of the spin-mixing parameter $b^2$. 
Furthermore, in Sec.~\ref{RashbaSec4} we analyze the effects of strain on the single-layer phosphorene placed in a perpendicular electric field using the effective spin-orbit Hamiltonian model and Dyakonov-Perel spin relaxation times. Finally, in Sec.~\ref{conclusions} the most important conclusions were summarized.
\section{Atomic structure and first principles calculation details}\label{DFTbasics}
\begin{figure}[t]
    \centering
    \includegraphics[width=1\columnwidth]{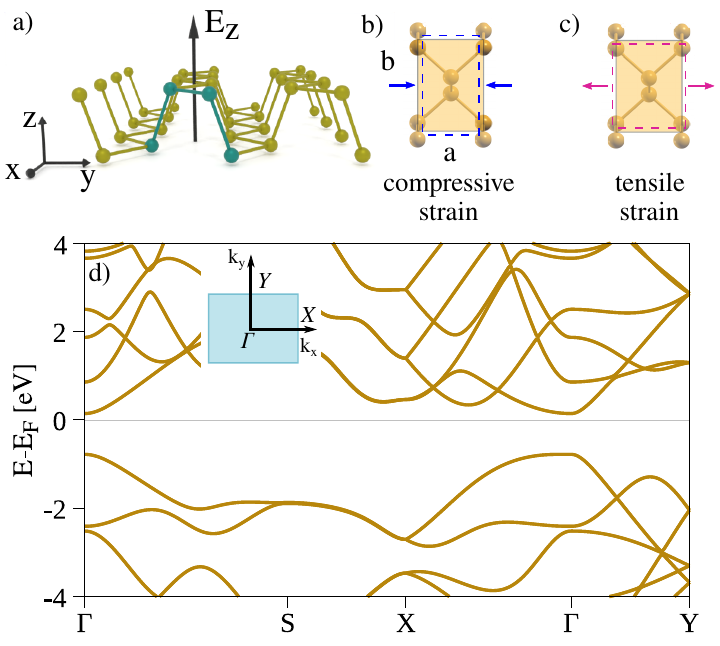}
    \caption{a)~A sketch of monolayer phosphorene with applied transverse electric field E$_\text{z}$. Atoms colored  green form the atomic basis in the unit cell; b) Top view of the unit cell of phosphorene (shaded yellow) under compressive strain indicated by blue arrows. The dashed line shows the response of the unit cell to the applied strain; c) same as in b) but for tensile strain. d)~Electronic band structure of phosphorene plotted long high symmetry lines of the first Brillouin zone shown in the inset. }
    \label{fig:structure}
\end{figure}

In Fig.~\ref{fig:structure}~a)-c), we show perspective and top views of the atomic structure models of the phosphorene monolayer (ML) subjected to strain. Under compressive strain the unit cell is squeezed in the direction parallel to strain, which induces non-zero pressure in the perpendicular direction. Relaxation of this uncompensated force leads to extending the structure the direction opposite to strain, as indicated by dashed blue line in Fig.~\ref{fig:structure}~b).

First-principles calculations
were performed using the plane wave Q{\sc{uantum}} ESPRESSO (QE) package~\cite{QE-2009,Giannozzi_2017}. We used the Perdew–Burke–Ernzerhof exchange-correlation functional ~\cite{perdew1996} as implemented in the full-relativistic SG15 optimized norm-conserving Vanderbilt (ONCV) pseudopotentials~\cite{Hamman2017,Scherpelz2016}.
The kinetic energy cut-offs 70\,Ry and 280\,Ry for the wave function and charge density, respectively were found to give well converged basis for the orbital and spin quantities. 
Periodic images of the crystal were separated by 20\,${\AA}$ vacuum in the $z$-direction. When electric field was applied, the vacuum was increased to 22\,${\AA}$ and  the dipole correction~\cite{B99} was taken into account to properly determine the energy offset due to dipole electric field effects.

The compressive and tensile strain was applied separately along two directions: zigzag\,(x) and armchair\,(y) by fixing the lattice vectors $\mathbf{b}$ and $\mathbf{a}$, respectively. Given the strain value, stress along the perpendicular direction was minimized by relaxing the second lattice vector using the variable-cell relaxation scheme. All structural calculations were done with $8\times 6\times 1$ Monkhorst-Pack \cite{MPack} $k$-point mesh. In actual calculations of electronic properties, the k-point mesh was increased to $16\times 12\times 1$. 
Optimized lattice parameters and corresponding pressure and total energy for each strain value are shown in Figs. S1 and S2 in the Supplementary Material.

\section{Strain effects on the spin-mixing in phosphorene}\label{b2parameter}
We start with investigating  effects of external strain on intrinsic SOC in phosphorene. 
In the following, we study the Elliott spin mixing parameter $b^2$ \cite{EY1954}, which provides comprehensive description of SOC in centrosymmetric non-magnetic materials.  It can be defined as follows. Without SOC, the Bloch states of a band $n$ corresponding to two spin states, $\psi_{n,\mathbf{k}}^\uparrow(\mathbf{r}) = a_{n,\mathbf{k}} (\mathbf{r})|\downarrow\rangle \exp(i \mathbf{k}\cdot \mathbf{r}) $ and $\psi_{n,\mathbf{k}}^\downarrow(\mathbf{r})=a_{n,\mathbf{k}} (\mathbf{r})|\downarrow\rangle \exp(i \mathbf{k}\cdot \mathbf{r})$, where $a_{n,\mathbf{k}}$ is a lattice periodic function, while $|\uparrow\rangle$,$|\downarrow\rangle$  are eigenstates of the $s_z$ spin one-half operator with eigenvalues $\pm \hbar /2$, representing pure spin \textit{up} ($\uparrow$) and spin \textit{down} ($\downarrow$) spinors, respectively. These states are degenerate at any $k$ due to time reversal and space inversion symmetry. Upon inclusion of SOC they become two component spinors
$\psi^\Uparrow_{n,\mathbf{k}}(\mathbf{r}) =[ a_{n,\mathbf{k}}(\mathbf{r}) |\uparrow\rangle + b_{n,\mathbf{k}} (\mathbf{r})|\downarrow\rangle]\exp(i \mathbf{k}\cdot \mathbf{r})$, 
 $\psi^\Downarrow_{n,\mathbf{k}}(\mathbf{r}) =[ a^*_{n,\mathbf{-k}}(\mathbf{r}) |\uparrow\rangle - b^*_{n,\mathbf{-k}} (\mathbf{r})|\downarrow\rangle]\exp(i \mathbf{k}\cdot \mathbf{r})$, 
 where $b_{n,\mathbf{k}} (\mathbf{r})$ is the spin component admixed by SOC. Typically  $|b_{n,\mathbf{k}}(\mathbf{r})| \ll |a_{n,\mathbf{k}}(\mathbf{r})|$ due to weakness of SOC, and $\psi^\Uparrow_{n,\mathbf{k}}(\mathbf{r})$, $\psi^\Downarrow_{n,\mathbf{k}}(\mathbf{r})$ still represent the majority spin up and majority spin down states, respectively.  
 Using first order perturbation theory one can show, that the admixture amplitude $b_{n,\mathbf{k}}(\mathbf{r}) $ is proportional to the strength of SOC in a band, $b _{n,\mathbf{k}} \sim \lambda_{\text{so}}$. 
 
 The spin mixing parameter can be defined as 
 $b^2_{n,\mathbf{k}} = \int_{BZ} |b_{n,\mathbf{k}}(\mathbf{r})|^2 d\mathbf{r}$, where the integration runs over the entire first Brillouin zone.  A non-zero $b^2_{n,\mathbf{k}}$ implies  $\vert \langle \psi^{\sigma}_{n,k} |\hat{s}_{z }|\psi^{\sigma}_{n,k}\rangle 
\vert < 1/2$, $\tilde{\sigma=\lbrace \Uparrow,\Downarrow\rbrace}$, since  $\psi^\Uparrow_{n,\mathbf{k}}(\mathbf{r})$ and $\psi^\Downarrow_{n,\mathbf{k}}(\mathbf{r})$ are no more the eigenstates of $\hat{s_z}$. This leads to the following definition of the spin mixing parameter \cite{Zimmermann} $b^2_{n,\mathbf{k} } = 1/2 - |\langle \psi^{\sigma}_{n,k} |\hat{s}_{z}|\psi^{\sigma}_{n,k}\rangle |/\hbar$.
For pure \textit{up} and \textit{down} spinors $b^2_{n,\mathbf{k}}=0$ and  $b^2_{n,\mathbf{k}}=0.5$ for fully spin mixed states. 

In the above definition we used as the spin basis the eigenstates of $\hat{s}_z$ operator, which set the spin quantization axis (SQA) along the $z$ axis (out-of-plane). However, one can choose the SQA arbitrarily and study the anisotropy of $b^2$. In our calculations we set SQA to be aligned with one of the Cartesian system axes, $\text{sqa}=\lbrace x,y,z\rbrace$, and  $\hat{s}_{z}$ operator in the definition of $b^2_{n,\mathbf{k}}=0$ represents the spin one-half operator diagonal in the spin basis given by the choice of \textit{sqa}.

Phosphorus is the second element of the group fourteen of periodic table with the atomic number Z=15. In elemental 2D materials the strength of the intrinsic SOC  $\lambda_{\text{so}}$ is proportional to $Z^2$,  $\lambda_{\text{so}} \sim Z^2$, suggesting that the relativistic effects on phosphorene are weak \cite{KFGF19}. Indeed, for unstrained  black phosphorus monolayer $b^2$ varies from  $10^{-5}$ to $10^{-4}$  \cite{KGF16,KFGF19}, placing phosphorene between graphene and silicene. 
Here we examine, how external strain modifies these values and the anisotropy of $b^2$. Because the electronic and mechanical properties of phosphorene are anisotropic, one can expect a similar anisotropic response of intrinsic SOC to external strain. 

\begin{figure}[t]
    \centering
\includegraphics[width=0.99\columnwidth]{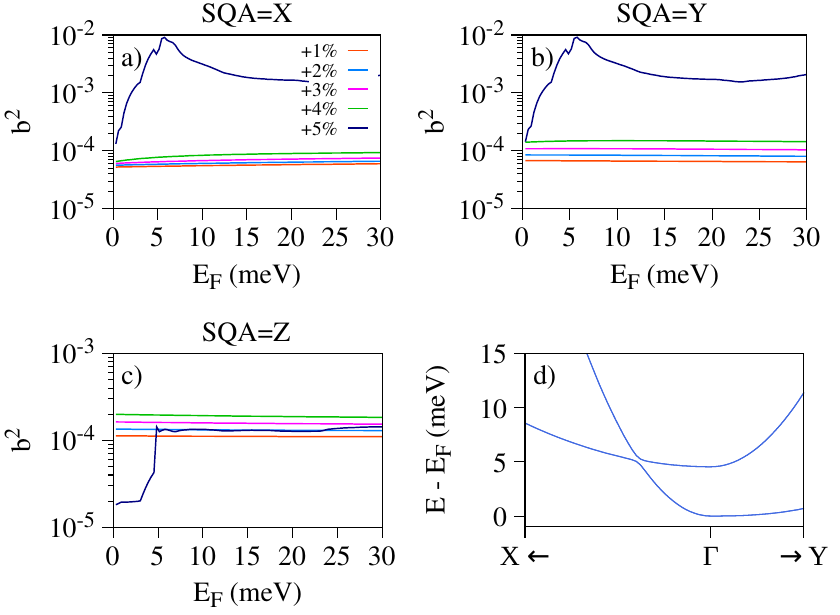}
    \caption{a)-c)~Average values of spin-mixing parameter $b^2$ of phosphorene electrons for each tensile strain applied along zigzag\,(x) direction. d)~Conduction bands interaction around $\Gamma$ point for strain +5\%. The kink for strain +5\% appears at 4.514\,meV and 5.725\,meV.  } 
    \label{fig:b2k}
\end{figure}

In Fig.~\ref{fig:b2k} we show the average spin mixing parameter $b^2$ in the conduction band plotted versus the Fermi energy for the tensile strain along the zigzag direction ($\Gamma X$ in the reciprocal space). For strain up to 4\,\% $b^2$ is almost insensitive to the applied strain. Within the considered doping range it takes the values close to  $10^{-4}$ and shows constant in-plane to out-of-plane anisotropy $b^2_\parallel / b^2_\perp \sim 2$. 
It can be understood by tracing the effects of tensile strain on the band structure. The main effect of the applied strain along the zigzag edge is the down-shift of the second conduction band (Fig. S11 in the Supplementary Material), whose contribution to $b^2$ for unstrained phosphorene is small. For 4\.\% strain $b^2$ is roughly twice the value for 1\,\%, independent on the chosen spin quantization axis (SQA). 

When strain grows above 4.8\,\%  the second conduction bands gets almost aligned with the first one shifting the anti-crossing toward the $\Gamma$ point (see Fig. \ref{fig:structure} d)).  Stretching phosphorene by 5\,\% places this anti-crossing approximately 6\,meV above the conduction band edge, see Fig. \ref{fig:b2k} d), elevating $b^2$ up to $10^{-2}$ for in-plane SQAs,  Fig. \ref{fig:b2k} a)-b),  and creating a spin hot spot \cite{fabian_spin_1998}. 
For SQA=Z $b^2$ starts from $2\cdot 10^{-5}$ close to the conduction band edge,  and jumps to $10^{-4}$ when the second conduction band enters the Fermi energy window ($E_F \le 5\,$meV), resulting in boosting and reversing spin mixing anisotropy: $b^2_\parallel / b^2_\perp \sim 10^{-2}$. 

The qualitative and quantitative change in spin mixing anisotropy will affect the Elliott-Yafet spin lifetimes $\tau_s$ in phosphorene. Recent spin injection experiments report $\tau_{s,\perp}/\tau_{s,\parallel} \sim 10$ \cite{CLT+24}, pointing to g-factor anisotropy as the responsible mechanism, since for unstrained monolayer and bulk black phosphorus $b^2$ anisotropy is opposite \cite{KGF16,ATK+17}. Here we show, that strain can be another factor affecting spin physics in phosphorene. 
Our results suggest that if strain is present, g-factor and spin mixing anisotropy of conduction electrons coincide and one can expect a giant spin lifetime anisotropy.

In contrast,  the valence band tensile strain along the zigzag direction leaves $b^2$ almost untouched. The relative change in $b^2$ for strain between 1\,\% and 5\,\% is up to  30\,\% (see Fig. S5 in the Supplemental Material). 

Strain applied along the armchair edge leaves the band structure and $b^2$ almost unaffected (Fig. S11 in the Supplementary Material) . The most pronounced effect is a systematic shift of the peak in $b^2$ for the valence band toward higher doping energies, Fig. S6 f)-g). The peak originates coincides with the valence band maximum, which in phosphorene is  located away from the $\Gamma$ point.

The PBE band gap of phosphorene 0.91\,eV is significantly smaller than the experimental value 2\,eV \cite{LWL+14}. Therefore, we have  examined how the correction of the phosphorene band gap affects the above results by performing calculations using  hybrid Heyd-Scuseria-Ernzerhof\,(HSE) exchange-correlation potential \cite{HSE}. The calculated HSE band structure and spin mixing parameter are shown in Fig. S10 in the Supplementary Material. 
We do not observe noticeable differences in $b^2$ between PBE and HSE cases except for a valence band near the X point. However, this region lies almost 1\,eV below the band maximum and is beyond the typical experimental doping range. All other minor discrepancies between the PBE and HSE cases do not change the overall order of magnitude of the parameter $b^2$. 

\section{Rashba effect in strained phosphorene}\label{RashbaSec4}

When the electric field in the $z$-direction is applied, the symmetry of a system
reduces to ${\bf C}_{2{\rm v}}$. The reduced symmetry additionally leads to the spin splitting of phosphorene bands, whose spin texture and the corresponding band splitting around the $\Gamma$ point can be modeled using the effective spin-orbit Hamiltonian 
\begin{eqnarray}\label{effmodel}
    H_{\rm SOC}^{\rm eff}&=& \lambda_1 k_x \sigma_y + \lambda_2 k_y \sigma_x+\lambda_3 k_x^3 \sigma_y+
    \lambda_4 k_y^3 \sigma_x\nonumber\\
    && + \lambda_5 k_xk_y^2 \sigma_y
    + \lambda_6 k_x^2k_y \sigma_x,
\end{eqnarray}
up to the terms qubic in momenta. To justify the model derived from the theory of invariants,
we mention that the two-fold rotation axis around the z direction changes a pair $(k_x,k_y)$
to $-(k_x,k_y)$, while the pseudovector ${\bm \sigma}=(\sigma_x,\sigma_y, \sigma_z)$
to $(-\sigma_x,-\sigma_y, \sigma_z)$. On the other hand, the nonsymmorphic group element, whose orthogonal part corresponds to the vertical mirror plane symmetry coinciding with the xz plane, transforms $(k_x,k_y)$ to $(k_x,-k_y)$ and $(\sigma_x,\sigma_y, \sigma_z)$ to 
$(-\sigma_x,\sigma_y,-\sigma_z)$. Using the above-mentioned transformation rules of the momentum and spin operators, up to the terms cubic in momenta the spin-orbit Hamiltonian can be written in a form given in~\eqref{effmodel}. The parameters $\lambda_{i}$, $i=1,...,6$, should be determined by fitting the model SOC Hamiltonian to the DFT data, having the range $k_{x/y}\in(0,\kappa_{x/y})$, where $\kappa_{x/y}=0.095/0.068\,\AA^{-1}$  corresponds to the $5\%$ of the $\Gamma X/\Gamma Y$ path in the case of the unstrained phosphorene. As the test case, we have analyzed the zero strain situation first and obtained anisotropic spin splitting of electrons and holes that is consistent with~\cite{PKS15}; the symmetric model of electron and holes discussed in~\cite{PFG+24} is not valid in our case.

\begin{figure*}[t]
    \centering
    \includegraphics[angle=270, width=0.99\textwidth]{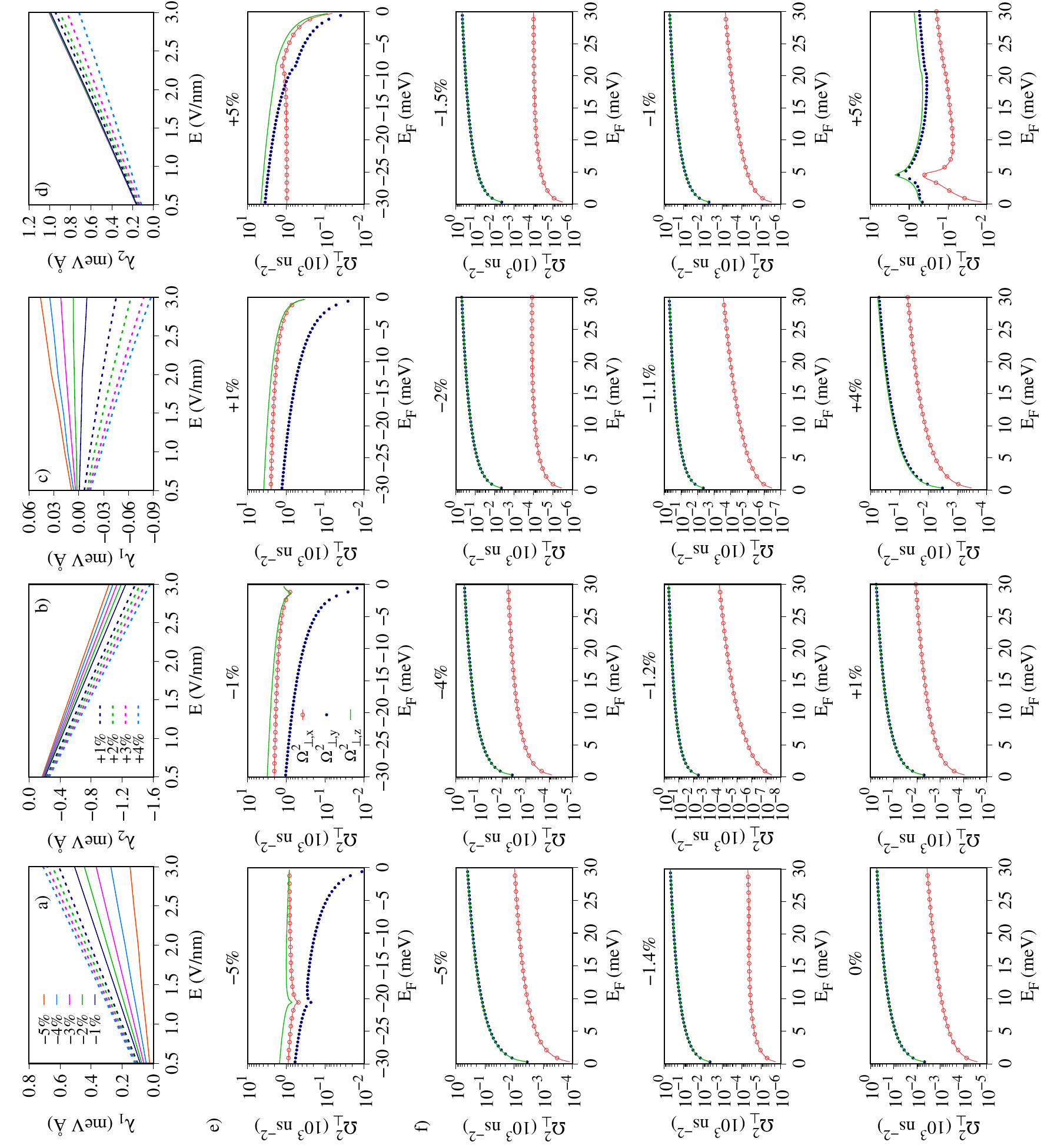}
    \caption{{\it Influence of electric field on spin physics of  phosphorene strained in the zigzag direction}. Parameters of linear terms $\lambda_{1}$, $\lambda_{2}$ for valence~a), b) and conduction band~c), d) are obtained after fitting the DFT data to the Hamiltonian model given in~\eqref{effmodel}. Additionally, for the applied electric field of E=1V/nm and different values of strain, we present the dependence of squared spin-orbit field components $\Omega_{\perp,i}^2$, $i = \{x, y, z\}$, for phosphorene hole e) and electron f) states, against the position of the Fermi level $E_{\rm F}$ measured from the top/bottom of the valence/conduction band.}
    \label{fig:DP_time}
\end{figure*}

Furthermore, assuming the electric field strengths from 0.5\,V/nm to 3\,V/nm, we have fitted the spin splitting energies and spin expectation values of the top valence and bottom conduction band close to the $\Gamma$ point. 
Our results indicate (as an illustration, see FIG.~S8 of the Supplemental Material) that the terms linear in momenta represent the dominant contribution to the overall spin-orbit field, whereas the terms cubic in momenta represent a correction. Thus, an overall picture of the spin texture of phosphorene electrons and holes can be obtained by analyzing the terms linear in momenta solely. In Fig. 3, we present the dependence of the parameters $\lambda_1$ and $\lambda_2$ on the electric field for different values of strain in the zigzag direction (results for strain in the armchair direction can be found in Supplemental Material, see FIG.~S9). 
In the case of valence bands, both $\lambda_1$ and $\lambda_2$ are negative and of the same order of magnitude, whereas in the case of the conduction bands (note that we have not analyzed the case of $+5\%$ strain in terms of the effective model due to the pronounced crossing of the conduction bands in the vicinity of the $\Gamma$ point, leading to the strong spin-mixing effect that can not be described in terms of the simple model~\eqref{effmodel}) the spin texture is dominated by the $\lambda_2 k_y \sigma_x$ term, suggesting that spins point mostly in the $x$ direction. Moreover, in Fig.~\ref{fig:DP_time} c) we see the parameter $\lambda_1$ changes its sign from positive to negative for compressive strain between 2 and 1 percent independently of the strength of the electric field. Thus, under appropriate strain, it is possible to obtain fully polarized spins in the x-direction. In order to confirm that, we compare the spin texture obtained using the effective model (up to the linear term) with the spin texture obtained with first-principles calculations. The spin texture obtained from DFT calculation indicates that the most  polarized spin texture appears for strain values close to -1\%. The discrepancy between the linear SOC model and the first-principles calculations can be explained by the presence of the SOC terms cubit in momenta ($k_x^3\sigma_y$) that are present even in the case of zero term $\propto  k_x \sigma_y$.

In the linear approximation, the resulting spin texture can be interpreted also in terms of the Dresselhaus and Rashba spin-orbit Hamiltonians, equal to $H_{\rm D}=\alpha_{\rm D} (k_y\sigma_x+k_x \sigma_y)$ and $H_{\rm R}=\alpha_{\rm R} (k_y\sigma_x-k_x \sigma_y)$, respectively. When comparing to the linear terms in~\eqref{effmodel}, we get $\alpha_{\rm D/R}=(\lambda_2\pm\lambda_1)/2$. In valence band case,
Rashba SOC is affected by strain roughly 2 times when changing the strain value from negative to positive, whereas the Dresselhaus SOC is weakly influenced by strain. Thus, for negative strain both the Rashba and Dresselhaus have comparable influence to the overall  spin texture, whereas for positive strain spoin texture is Rashba SOC dominated.  In the case of conduction bands, due to $|\lambda_1|\ll|\lambda_2|$, both Rashba and Dresselhaus SOC are almost identical, converging to the condition for permanent spin texture $\alpha_{\rm R}=\alpha_{\rm D}$. Since $\alpha_{\rm D}$ decreases with increasing strain from negative to positive values for $70\%$, whereas the Rashba SOC strenght is barely affected by strain, the anisotropy of the in-plane spin texture can be highly tunable by strain, which will be confirmed by the Dyakonov-Perel spin relaxation calculations in Sec.~\ref{spinrelaxation}.

\subsection{Rashba field induced spin relaxation in strained phosphorene}\label{spinrelaxation}

Spin relaxation times are an experimentally measurable quantity influenced by the strength of spin-orbit coupling. Without the electric field, spin relaxation times of phosphorene~\cite{ATK+17,CLT+24}  are determined by the Elliott-Yafet mechanism~\cite{EY1954}. However, when an electric field is applied, the Dyakonov-Perel (DP) spin relaxation mechanism~\cite{dyakonov_1971R} is activated. In the DP regime, the spin relaxation time $\tau_{i}$ can be computed using the following relation
\begin{equation}\label{spinrelformula}
    \tau_{i}^{-1}=\Omega_{\perp,i}^2\tau_p,
\end{equation}
where $\Omega_{\perp,i}^2 = \langle \Omega^2\rangle - \langle \Omega^2_i\rangle$ corresponds to the Fermi surface average of the squared spin-orbit field component $\Omega_{{\bf k}\perp,i}^2$ that is perpendicular to the spin orientation $i=\{x,y,z\}$, while $\tau_p$ is the momentum relaxation time. From the first-principle calculations, one can directly extract the spin-orbit field $ \Omega_{\bf{k},i}$ for a spin-split  band at the given ${\bf k}$ point of the BZ using the relation
\begin{equation}
\Omega_{\mathbf{k},i}=\frac{\Delta_{\rm so}}{\hbar}\frac{s_i}{|s|},
\end{equation}
in which the $\Delta_{\rm so}$ corresponds to the spin splitting value, while $s_i$ is equal to the expectation value of the spin one-half operator at ${\bf k}$. Since $\tau_p$ in (\ref{spinrelformula}) depends on sample quality, we set $\tau_p=1$ and discuss below only the effects of $\Omega_{i}$.

For the strain in the zigzag direction and $E=1$ V/nm, results of such calculations for the top valence and bottom conduction band are presented in Fig.~\ref{fig:DP_time} e) and Fig.~\ref{fig:DP_time} f), respectively.
Fig.~\ref{fig:DP_time} e) shows that spin relaxation rates of phosphorene  holes  are not changed significantly by varying strain from $-5\%$ to $5\%$. On the other hand, strain significantly changes the spin relaxation rates of phosphorene electrons, and in particular spin relaxation anisotropy. 
Our calculations show that for negative strain between $-2\%$ and $-1\%$ there is significant magnitude decrease of spin relaxation rate $\Omega_{\perp,x}^2$, consistent with the minimized value of the parameter $\lambda_1$ in this regime. Specifically, for the strain value of $-1.2\%$ there is a three-order-of-magnitude decrease of $\Omega_{\perp,x}^2$ with respect to the zero strain case, suggesting the regime of almost permanent spin texture. The corresponding spin relaxation anisotropy (assuming isotropic momentum scattering) is $\Omega_{\perp,x}^2/\Omega_{\perp,y/z}^2\approx 10^{5}$ and suggest almost scattering-free propagation of spins polarized along the zigzag direction.  Thus, we have shown that compressive strain between $-1.5\%$ and $-1\%$ has a giant effect on the spin relaxation anisotropy of phosphorene electrons.  
The overall change in D-P spin relaxation anisotropy for the considered strain values in the zigzag direction is  $\approx 10^{3}$, and places strain as a very effective knob for tuning spin anisotropy in phosphorene. 

When analyzing the effect of positive strain (up to 4.5\%)
on spin-relaxation rates, we notice that spin-orbit field component $\Omega_{\perp,x}^2$ increase due to the increase of the parameter $\lambda_1$, whereas the small decrease of the $\lambda_2$ parameter correspond to decrease of the  $\Omega_{\perp,y/z}^2$ spin-orbit field components. Finally, the giant change of all spin-relaxation rates for $+5\%$ stretch is a consequence of the anti-crossing  entering the Fermi energy window analyzed.

In Fig.~S9 of the Supplementary Material, we analyze the influence of the electric field $E=1$ V/nm on the spin-orbit field components, assuming the strain in the armchair direction. Our results indicate that strain in the armchair direction has negligible effect on the spin physics of phosphorene holes and the spin-orbit fields of phosphorene electrons $\Omega_{\perp,y/z}^2$, maximally affecting the spin-orbit field $\Omega_{\perp,x}^2$ for a factor of 5.


\section{Conclusions}\label{conclusions}
We studied the effects of strain on the spin physics of monolayer phosphorene in the presence of (non)zero electric field in the direction perpendicular to the studied monolayer. In the zero electric field case, bands of phosphorene are degenerate due to the presence of the inversion symmetry and the dominant spin relaxation times is of the Elliott-Yafet type. Focusing on the top valence and bottom conduction bands, we have demonstrated small and continuous changes of the Elliott-Yafet spin relaxation times under strain, if there is no band anti-crossing appearing at the given doping energy. Otherwise, due to strain-induced spin hot spot a significant change of Elliott-Yafet spin relaxation takes place resulting in reversed and amplified spin lifetime anisotropy. In this case, due to the synergy of spin mixing and g-factor anisotropy, it is expected that a giant anisotropy of spin lifetime will be observed in spin injection experiments.

Furthermore, we investigated the strain-induced changes in the spin texture of electrons and holes of phosphorene subjected to the presence of perpendicular electric field. To this end we employed the effective spin-orbit Hamiltonian modeling and the numerical calculation of spin relaxation rates triggered by the  Dyakonov-Perel spin relaxation
mechanism.  Whereas for the valence bands strain effects are minimal, in the case of the conduction bands we showed that strain triggers a significant change of spin relaxation rate and its anisotropy.  By varying strain from -5\,\% to 5\,\% spin lifetime anisotropy changes by the factor of $10^3$. Thus strain can be successfully used for tuning the spin texture and spin anisotropy of phosphorene electrons. 

\acknowledgments
P. J. and M.K.~acknowledge  financial support provided by the National Center for Research and Development (NCBR) under the V4-Japan project BGapEng V4-JAPAN/2/46/BGapEng/2022 and support from the Interdisciplinary Centre for Mathematical and Computational Modelling (ICM), University of Warsaw (UW), within grant no. G83-27. 
P.J. acknowledges financial support from the National Agency for Academic Exchange under the STER program – Internationalization of Doctoral Schools, project: International from the beginning – wsparcie umi\k{e}dzynarodowienia.
M.M. acknowledges the financial support
provided by the Ministry of Education, Science, and Technological Development of the Republic of Serbia. This project has received funding from the European Union's Horizon 2020 Research and Innovation Programme under the Programme SASPRO 2 COFUND Marie Sklodowska-Curie grant agreement No. 945478.


\bibliography{bibliography}

\begin{thebibliography}{41}%
\makeatletter
\providecommand \@ifxundefined [1]{%
 \@ifx{#1\undefined}
}%
\providecommand \@ifnum [1]{%
 \ifnum #1\expandafter \@firstoftwo
 \else \expandafter \@secondoftwo
 \fi
}%
\providecommand \@ifx [1]{%
 \ifx #1\expandafter \@firstoftwo
 \else \expandafter \@secondoftwo
 \fi
}%
\providecommand \natexlab [1]{#1}%
\providecommand \enquote  [1]{``#1''}%
\providecommand \bibnamefont  [1]{#1}%
\providecommand \bibfnamefont [1]{#1}%
\providecommand \citenamefont [1]{#1}%
\providecommand \href@noop [0]{\@secondoftwo}%
\providecommand \href [0]{\begingroup \@sanitize@url \@href}%
\providecommand \@href[1]{\@@startlink{#1}\@@href}%
\providecommand \@@href[1]{\endgroup#1\@@endlink}%
\providecommand \@sanitize@url [0]{\catcode `\\12\catcode `\$12\catcode `\&12\catcode `\#12\catcode `\^12\catcode `\_12\catcode `\%12\relax}%
\providecommand \@@startlink[1]{}%
\providecommand \@@endlink[0]{}%
\providecommand \url  [0]{\begingroup\@sanitize@url \@url }%
\providecommand \@url [1]{\endgroup\@href {#1}{\urlprefix }}%
\providecommand \urlprefix  [0]{URL }%
\providecommand \Eprint [0]{\href }%
\providecommand \doibase [0]{https://doi.org/}%
\providecommand \selectlanguage [0]{\@gobble}%
\providecommand \bibinfo  [0]{\@secondoftwo}%
\providecommand \bibfield  [0]{\@secondoftwo}%
\providecommand \translation [1]{[#1]}%
\providecommand \BibitemOpen [0]{}%
\providecommand \bibitemStop [0]{}%
\providecommand \bibitemNoStop [0]{.\EOS\space}%
\providecommand \EOS [0]{\spacefactor3000\relax}%
\providecommand \BibitemShut  [1]{\csname bibitem#1\endcsname}%
\let\auto@bib@innerbib\@empty
\bibitem [{\citenamefont {Wolf}\ \emph {et~al.}(2001)\citenamefont {Wolf}, \citenamefont {Awschalom}, \citenamefont {Buhrman}, \citenamefont {Daughton}, \citenamefont {von Molnár}, \citenamefont {Roukes}, \citenamefont {Chtchelkanova},\ and\ \citenamefont {Treger}}]{W+01}%
  \BibitemOpen
  \bibfield  {author} {\bibinfo {author} {\bibfnamefont {S.~A.}\ \bibnamefont {Wolf}}, \bibinfo {author} {\bibfnamefont {D.~D.}\ \bibnamefont {Awschalom}}, \bibinfo {author} {\bibfnamefont {R.~A.}\ \bibnamefont {Buhrman}}, \bibinfo {author} {\bibfnamefont {J.~M.}\ \bibnamefont {Daughton}}, \bibinfo {author} {\bibfnamefont {S.}~\bibnamefont {von Molnár}}, \bibinfo {author} {\bibfnamefont {M.~L.}\ \bibnamefont {Roukes}}, \bibinfo {author} {\bibfnamefont {A.~Y.}\ \bibnamefont {Chtchelkanova}},\ and\ \bibinfo {author} {\bibfnamefont {D.~M.}\ \bibnamefont {Treger}},\ }\bibfield  {title} {\bibinfo {title} {Spintronics: A spin-based electronics vision for the future},\ }\href {https://doi.org/10.1126/science.1065389} {\bibfield  {journal} {\bibinfo  {journal} {Science}\ }\textbf {\bibinfo {volume} {294}},\ \bibinfo {pages} {1488} (\bibinfo {year} {2001})}\BibitemShut {NoStop}%
\bibitem [{\citenamefont {\ifmmode \check{Z}\else \v{Z}\fi{}uti\ifmmode~\acute{c}\else \'{c}\fi{}}\ \emph {et~al.}(2004)\citenamefont {\ifmmode \check{Z}\else \v{Z}\fi{}uti\ifmmode~\acute{c}\else \'{c}\fi{}}, \citenamefont {Fabian},\ and\ \citenamefont {Das~Sarma}}]{ZFS04}%
  \BibitemOpen
  \bibfield  {author} {\bibinfo {author} {\bibfnamefont {I.}~\bibnamefont {\ifmmode \check{Z}\else \v{Z}\fi{}uti\ifmmode~\acute{c}\else \'{c}\fi{}}}, \bibinfo {author} {\bibfnamefont {J.}~\bibnamefont {Fabian}},\ and\ \bibinfo {author} {\bibfnamefont {S.}~\bibnamefont {Das~Sarma}},\ }\bibfield  {title} {\bibinfo {title} {Spintronics: Fundamentals and applications},\ }\href {https://doi.org/10.1103/RevModPhys.76.323} {\bibfield  {journal} {\bibinfo  {journal} {Rev. Mod. Phys.}\ }\textbf {\bibinfo {volume} {76}},\ \bibinfo {pages} {323} (\bibinfo {year} {2004})}\BibitemShut {NoStop}%
\bibitem [{\citenamefont {Fabian}\ \emph {et~al.}(2007)\citenamefont {Fabian}, \citenamefont {Matos-Abiague}, \citenamefont {Ertler}, \citenamefont {Stano},\ and\ \citenamefont {\v{Z}uti\'c}}]{FME+04}%
  \BibitemOpen
  \bibfield  {author} {\bibinfo {author} {\bibfnamefont {J.}~\bibnamefont {Fabian}}, \bibinfo {author} {\bibfnamefont {A.}~\bibnamefont {Matos-Abiague}}, \bibinfo {author} {\bibfnamefont {C.}~\bibnamefont {Ertler}}, \bibinfo {author} {\bibfnamefont {P.}~\bibnamefont {Stano}},\ and\ \bibinfo {author} {\bibfnamefont {I.}~\bibnamefont {\v{Z}uti\'c}},\ }\bibfield  {title} {\bibinfo {title} {Semiconductor spintronics},\ }\href {https://doi.org/http://www.physics.sk/aps/pub.php?y=2007&pub=aps-07-04} {\bibfield  {journal} {\bibinfo  {journal} {Acta Phys. Slovaca}\ }\textbf {\bibinfo {volume} {57}},\ \bibinfo {pages} {565} (\bibinfo {year} {2007})}\BibitemShut {NoStop}%
\bibitem [{\citenamefont {Han}\ \emph {et~al.}(2014)\citenamefont {Han}, \citenamefont {Kawakami}, \citenamefont {Gmitra},\ and\ \citenamefont {Fabian}}]{HKG+14}%
  \BibitemOpen
  \bibfield  {author} {\bibinfo {author} {\bibfnamefont {W.}~\bibnamefont {Han}}, \bibinfo {author} {\bibfnamefont {R.~K.}\ \bibnamefont {Kawakami}}, \bibinfo {author} {\bibfnamefont {M.}~\bibnamefont {Gmitra}},\ and\ \bibinfo {author} {\bibfnamefont {J.}~\bibnamefont {Fabian}},\ }\bibfield  {title} {\bibinfo {title} {Graphene spintronics},\ }\href {https://doi.org/10.1038/nnano.2014.214} {\bibfield  {journal} {\bibinfo  {journal} {Nature Nanotechnology}\ }\textbf {\bibinfo {volume} {9}},\ \bibinfo {pages} {794} (\bibinfo {year} {2014})}\BibitemShut {NoStop}%
\bibitem [{\citenamefont {Jansen}(2012)}]{J12}%
  \BibitemOpen
  \bibfield  {author} {\bibinfo {author} {\bibfnamefont {R.}~\bibnamefont {Jansen}},\ }\bibfield  {title} {\bibinfo {title} {Silicon spintronics},\ }\href {https://doi.org/10.1038/nmat3293} {\bibfield  {journal} {\bibinfo  {journal} {Nature Materials}\ }\textbf {\bibinfo {volume} {11}},\ \bibinfo {pages} {400} (\bibinfo {year} {2012})}\BibitemShut {NoStop}%
\bibitem [{\citenamefont {Liu}\ \emph {et~al.}(2014)\citenamefont {Liu}, \citenamefont {Neal}, \citenamefont {Zhu}, \citenamefont {Luo}, \citenamefont {Xu}, \citenamefont {Tom\'{a}nek},\ and\ \citenamefont {Ye}}]{LNZ+14}%
  \BibitemOpen
  \bibfield  {author} {\bibinfo {author} {\bibfnamefont {H.}~\bibnamefont {Liu}}, \bibinfo {author} {\bibfnamefont {A.~T.}\ \bibnamefont {Neal}}, \bibinfo {author} {\bibfnamefont {Z.}~\bibnamefont {Zhu}}, \bibinfo {author} {\bibfnamefont {Z.}~\bibnamefont {Luo}}, \bibinfo {author} {\bibfnamefont {X.}~\bibnamefont {Xu}}, \bibinfo {author} {\bibfnamefont {D.}~\bibnamefont {Tom\'{a}nek}},\ and\ \bibinfo {author} {\bibfnamefont {P.~D.}\ \bibnamefont {Ye}},\ }\bibfield  {title} {\bibinfo {title} {Phosphorene: An unexplored 2d semiconductor with a high hole mobility},\ }\href {https://doi.org/10.1021/nn501226z} {\bibfield  {journal} {\bibinfo  {journal} {ACS Nano}\ }\textbf {\bibinfo {volume} {8}},\ \bibinfo {pages} {4033–4041} (\bibinfo {year} {2014})}\BibitemShut {NoStop}%
\bibitem [{\citenamefont {Lu}\ \emph {et~al.}(2014)\citenamefont {Lu}, \citenamefont {Nan}, \citenamefont {Hong}, \citenamefont {Chen}, \citenamefont {Zhu}, \citenamefont {Liang}, \citenamefont {Ma}, \citenamefont {Ni}, \citenamefont {Jin},\ and\ \citenamefont {Zhang}}]{LNH+14}%
  \BibitemOpen
  \bibfield  {author} {\bibinfo {author} {\bibfnamefont {W.}~\bibnamefont {Lu}}, \bibinfo {author} {\bibfnamefont {H.}~\bibnamefont {Nan}}, \bibinfo {author} {\bibfnamefont {J.}~\bibnamefont {Hong}}, \bibinfo {author} {\bibfnamefont {Y.}~\bibnamefont {Chen}}, \bibinfo {author} {\bibfnamefont {C.}~\bibnamefont {Zhu}}, \bibinfo {author} {\bibfnamefont {Z.}~\bibnamefont {Liang}}, \bibinfo {author} {\bibfnamefont {X.}~\bibnamefont {Ma}}, \bibinfo {author} {\bibfnamefont {Z.}~\bibnamefont {Ni}}, \bibinfo {author} {\bibfnamefont {C.}~\bibnamefont {Jin}},\ and\ \bibinfo {author} {\bibfnamefont {Z.}~\bibnamefont {Zhang}},\ }\bibfield  {title} {\bibinfo {title} {Plasma-assisted fabrication of monolayer phosphorene and its raman characterization},\ }\href {https://doi.org/10.1007/s12274-014-0446-7} {\bibfield  {journal} {\bibinfo  {journal} {Nano Research}\ }\textbf {\bibinfo {volume} {7}},\ \bibinfo {pages} {853} (\bibinfo {year} {2014})}\BibitemShut {NoStop}%
\bibitem [{\citenamefont {Liang}\ \emph {et~al.}(2014)\citenamefont {Liang}, \citenamefont {Wang}, \citenamefont {Lin}, \citenamefont {Sumpter}, \citenamefont {Meunier},\ and\ \citenamefont {Pan}}]{LWL+14}%
  \BibitemOpen
  \bibfield  {author} {\bibinfo {author} {\bibfnamefont {L.}~\bibnamefont {Liang}}, \bibinfo {author} {\bibfnamefont {J.}~\bibnamefont {Wang}}, \bibinfo {author} {\bibfnamefont {W.}~\bibnamefont {Lin}}, \bibinfo {author} {\bibfnamefont {B.~G.}\ \bibnamefont {Sumpter}}, \bibinfo {author} {\bibfnamefont {V.}~\bibnamefont {Meunier}},\ and\ \bibinfo {author} {\bibfnamefont {M.}~\bibnamefont {Pan}},\ }\bibfield  {title} {\bibinfo {title} {Electronic bandgap and edge reconstruction in phosphorene materials},\ }\href {https://doi.org/10.1021/nl502892t} {\bibfield  {journal} {\bibinfo  {journal} {Nano Lett.}\ }\textbf {\bibinfo {volume} {14}},\ \bibinfo {pages} {6400} (\bibinfo {year} {2014})}\BibitemShut {NoStop}%
\bibitem [{\citenamefont {Zhang}\ \emph {et~al.}(2014)\citenamefont {Zhang}, \citenamefont {Yang}, \citenamefont {Xu}, \citenamefont {Wang}, \citenamefont {Li}, \citenamefont {Ghufran}, \citenamefont {Zhang}, \citenamefont {Yu}, \citenamefont {Zhang}, \citenamefont {Qin},\ and\ \citenamefont {Lu}}]{ZYX+14}%
  \BibitemOpen
  \bibfield  {author} {\bibinfo {author} {\bibfnamefont {S.}~\bibnamefont {Zhang}}, \bibinfo {author} {\bibfnamefont {J.}~\bibnamefont {Yang}}, \bibinfo {author} {\bibfnamefont {R.}~\bibnamefont {Xu}}, \bibinfo {author} {\bibfnamefont {F.}~\bibnamefont {Wang}}, \bibinfo {author} {\bibfnamefont {W.}~\bibnamefont {Li}}, \bibinfo {author} {\bibfnamefont {M.}~\bibnamefont {Ghufran}}, \bibinfo {author} {\bibfnamefont {Y.-W.}\ \bibnamefont {Zhang}}, \bibinfo {author} {\bibfnamefont {Z.}~\bibnamefont {Yu}}, \bibinfo {author} {\bibfnamefont {G.}~\bibnamefont {Zhang}}, \bibinfo {author} {\bibfnamefont {Q.}~\bibnamefont {Qin}},\ and\ \bibinfo {author} {\bibfnamefont {Y.}~\bibnamefont {Lu}},\ }\bibfield  {title} {\bibinfo {title} {Extraordinary photoluminescence and strong temperature/angle-dependent raman responses in few-layer phosphorene},\ }\href {https://doi.org/10.1021/nn503893j} {\bibfield  {journal} {\bibinfo  {journal} {ACS Nano}\ }\textbf {\bibinfo {volume} {8}},\ \bibinfo {pages} {9590} (\bibinfo {year}
  {2014})},\ \bibinfo {note} {doi: 10.1021/nn503893j}\BibitemShut {NoStop}%
\bibitem [{\citenamefont {Castellanos-Gomez}\ \emph {et~al.}(2014)\citenamefont {Castellanos-Gomez}, \citenamefont {Vicarelli}, \citenamefont {Prada}, \citenamefont {Island}, \citenamefont {Narasimha-Acharya}, \citenamefont {Blanter}, \citenamefont {Groenendijk}, \citenamefont {Buscema}, \citenamefont {Steele}, \citenamefont {Alvarez}, \citenamefont {Zandbergen}, \citenamefont {Palacios},\ and\ \citenamefont {van~der Zant}}]{CVP+14}%
  \BibitemOpen
  \bibfield  {author} {\bibinfo {author} {\bibfnamefont {A.}~\bibnamefont {Castellanos-Gomez}}, \bibinfo {author} {\bibfnamefont {L.}~\bibnamefont {Vicarelli}}, \bibinfo {author} {\bibfnamefont {E.}~\bibnamefont {Prada}}, \bibinfo {author} {\bibfnamefont {J.~O.}\ \bibnamefont {Island}}, \bibinfo {author} {\bibfnamefont {K.~L.}\ \bibnamefont {Narasimha-Acharya}}, \bibinfo {author} {\bibfnamefont {S.~I.}\ \bibnamefont {Blanter}}, \bibinfo {author} {\bibfnamefont {D.~J.}\ \bibnamefont {Groenendijk}}, \bibinfo {author} {\bibfnamefont {M.}~\bibnamefont {Buscema}}, \bibinfo {author} {\bibfnamefont {G.~A.}\ \bibnamefont {Steele}}, \bibinfo {author} {\bibfnamefont {J.~V.}\ \bibnamefont {Alvarez}}, \bibinfo {author} {\bibfnamefont {H.~W.}\ \bibnamefont {Zandbergen}}, \bibinfo {author} {\bibfnamefont {J.~J.}\ \bibnamefont {Palacios}},\ and\ \bibinfo {author} {\bibfnamefont {H.~S.~J.}\ \bibnamefont {van~der Zant}},\ }\bibfield  {title} {\bibinfo {title} {Isolation and characterization of few-layer black phosphorus},\
  }\href {https://doi.org/10.1088/2053-1583/1/2/025001} {\bibfield  {journal} {\bibinfo  {journal} {2D Materials}\ }\textbf {\bibinfo {volume} {1}},\ \bibinfo {pages} {025001} (\bibinfo {year} {2014})}\BibitemShut {NoStop}%
\bibitem [{\citenamefont {Qiao}\ \emph {et~al.}(2014)\citenamefont {Qiao}, \citenamefont {Kong}, \citenamefont {Hu}, \citenamefont {Yang},\ and\ \citenamefont {Ji}}]{QKH+14}%
  \BibitemOpen
  \bibfield  {author} {\bibinfo {author} {\bibfnamefont {J.}~\bibnamefont {Qiao}}, \bibinfo {author} {\bibfnamefont {X.}~\bibnamefont {Kong}}, \bibinfo {author} {\bibfnamefont {Z.-X.}\ \bibnamefont {Hu}}, \bibinfo {author} {\bibfnamefont {F.}~\bibnamefont {Yang}},\ and\ \bibinfo {author} {\bibfnamefont {W.}~\bibnamefont {Ji}},\ }\bibfield  {title} {\bibinfo {title} {High-mobility transport anisotropy and linear dichroism in few-layer black phosphorus},\ }\href@noop {} {\bibfield  {journal} {\bibinfo  {journal} {Nat. Commun.}\ }\textbf {\bibinfo {volume} {5}},\ \bibinfo {pages} {4475} (\bibinfo {year} {2014})}\BibitemShut {NoStop}%
\bibitem [{\citenamefont {Avsar}\ \emph {et~al.}(2017)\citenamefont {Avsar}, \citenamefont {Tan}, \citenamefont {Kurpas}, \citenamefont {Gmitra}, \citenamefont {Watanabe}, \citenamefont {Taniguchi}, \citenamefont {Fabian},\ and\ \citenamefont {{\"{O}}zyilmaz}}]{ATK+17}%
  \BibitemOpen
  \bibfield  {author} {\bibinfo {author} {\bibfnamefont {A.}~\bibnamefont {Avsar}}, \bibinfo {author} {\bibfnamefont {J.~Y.}\ \bibnamefont {Tan}}, \bibinfo {author} {\bibfnamefont {M.}~\bibnamefont {Kurpas}}, \bibinfo {author} {\bibfnamefont {M.}~\bibnamefont {Gmitra}}, \bibinfo {author} {\bibfnamefont {K.}~\bibnamefont {Watanabe}}, \bibinfo {author} {\bibfnamefont {T.}~\bibnamefont {Taniguchi}}, \bibinfo {author} {\bibfnamefont {J.}~\bibnamefont {Fabian}},\ and\ \bibinfo {author} {\bibfnamefont {B.}~\bibnamefont {{\"{O}}zyilmaz}},\ }\bibfield  {title} {\bibinfo {title} {Gate-tunable black phosphorus spin valve with nanosecond spin lifetimes},\ }\href {https://doi.org/10.1038/nphys4141} {\bibfield  {journal} {\bibinfo  {journal} {Nat. Phys.}\ }\textbf {\bibinfo {volume} {13}},\ \bibinfo {pages} {888} (\bibinfo {year} {2017})}\BibitemShut {NoStop}%
\bibitem [{\citenamefont {Cording}\ \emph {et~al.}(2024)\citenamefont {Cording}, \citenamefont {Liu}, \citenamefont {Tan}, \citenamefont {Watanabe}, \citenamefont {Taniguchi}, \citenamefont {Avsar},\ and\ \citenamefont {Özyilmaz}}]{CLT+24}%
  \BibitemOpen
  \bibfield  {author} {\bibinfo {author} {\bibfnamefont {L.}~\bibnamefont {Cording}}, \bibinfo {author} {\bibfnamefont {J.}~\bibnamefont {Liu}}, \bibinfo {author} {\bibfnamefont {J.~Y.}\ \bibnamefont {Tan}}, \bibinfo {author} {\bibfnamefont {K.}~\bibnamefont {Watanabe}}, \bibinfo {author} {\bibfnamefont {T.}~\bibnamefont {Taniguchi}}, \bibinfo {author} {\bibfnamefont {A.}~\bibnamefont {Avsar}},\ and\ \bibinfo {author} {\bibfnamefont {B.}~\bibnamefont {Özyilmaz}},\ }\bibfield  {title} {\bibinfo {title} {Highly anisotropic spin transport in ultrathin black phosphorus},\ }\href {https://doi.org/10.1038/s41563-023-01779-8} {\bibfield  {journal} {\bibinfo  {journal} {Nature Materials}\ }\textbf {\bibinfo {volume} {23}},\ \bibinfo {pages} {479} (\bibinfo {year} {2024})}\BibitemShut {NoStop}%
\bibitem [{\citenamefont {Kurpas}\ \emph {et~al.}(2016)\citenamefont {Kurpas}, \citenamefont {Gmitra},\ and\ \citenamefont {Fabian}}]{KGF16}%
  \BibitemOpen
  \bibfield  {author} {\bibinfo {author} {\bibfnamefont {M.}~\bibnamefont {Kurpas}}, \bibinfo {author} {\bibfnamefont {M.}~\bibnamefont {Gmitra}},\ and\ \bibinfo {author} {\bibfnamefont {J.}~\bibnamefont {Fabian}},\ }\bibfield  {title} {\bibinfo {title} {Spin-orbit coupling and spin relaxation in phosphorene: Intrinsic versus extrinsic effects},\ }\href {https://doi.org/10.1103/PhysRevB.94.155423} {\bibfield  {journal} {\bibinfo  {journal} {Phys. Rev. B}\ }\textbf {\bibinfo {volume} {94}},\ \bibinfo {pages} {155423} (\bibinfo {year} {2016})}\BibitemShut {NoStop}%
\bibitem [{\citenamefont {Peralta}\ \emph {et~al.}(2024)\citenamefont {Peralta}, \citenamefont {Freire}, \citenamefont {Gonz\'alez-Hern\'andez},\ and\ \citenamefont {Mireles}}]{PFG+24}%
  \BibitemOpen
  \bibfield  {author} {\bibinfo {author} {\bibfnamefont {M.}~\bibnamefont {Peralta}}, \bibinfo {author} {\bibfnamefont {D.~A.}\ \bibnamefont {Freire}}, \bibinfo {author} {\bibfnamefont {R.}~\bibnamefont {Gonz\'alez-Hern\'andez}},\ and\ \bibinfo {author} {\bibfnamefont {F.}~\bibnamefont {Mireles}},\ }\bibfield  {title} {\bibinfo {title} {Spin-orbit coupling effects in single-layer phosphorene},\ }\href {https://doi.org/10.1103/PhysRevB.110.085404} {\bibfield  {journal} {\bibinfo  {journal} {Phys. Rev. B}\ }\textbf {\bibinfo {volume} {110}},\ \bibinfo {pages} {085404} (\bibinfo {year} {2024})}\BibitemShut {NoStop}%
\bibitem [{\citenamefont {Milivojevi{\' c}}\ \emph {et~al.}(2023)\citenamefont {Milivojevi{\' c}}, \citenamefont {Gmitra}, \citenamefont {Kurpas}, \citenamefont {{\v S}tich},\ and\ \citenamefont {Fabian}}]{MGK+23}%
  \BibitemOpen
  \bibfield  {author} {\bibinfo {author} {\bibfnamefont {M.}~\bibnamefont {Milivojevi{\' c}}}, \bibinfo {author} {\bibfnamefont {M.}~\bibnamefont {Gmitra}}, \bibinfo {author} {\bibfnamefont {M.}~\bibnamefont {Kurpas}}, \bibinfo {author} {\bibfnamefont {I.}~\bibnamefont {{\v S}tich}},\ and\ \bibinfo {author} {\bibfnamefont {J.}~\bibnamefont {Fabian}},\ }\bibfield  {title} {\bibinfo {title} {Proximity-induced spin-orbit coupling in phosphorene on a wse2 monolayer},\ }\href@noop {} {\bibfield  {journal} {\bibinfo  {journal} {Phys. Rev. B}\ }\textbf {\bibinfo {volume} {108}},\ \bibinfo {pages} {115311} (\bibinfo {year} {2023})}\BibitemShut {NoStop}%
\bibitem [{\citenamefont {Milivojevi{\' c}}\ \emph {et~al.}(2024)\citenamefont {Milivojevi{\' c}}, \citenamefont {Gmitra}, \citenamefont {Kurpas}, \citenamefont {{\v S}tich},\ and\ \citenamefont {Fabian}}]{MGK+24}%
  \BibitemOpen
  \bibfield  {author} {\bibinfo {author} {\bibfnamefont {M.}~\bibnamefont {Milivojevi{\' c}}}, \bibinfo {author} {\bibfnamefont {M.}~\bibnamefont {Gmitra}}, \bibinfo {author} {\bibfnamefont {M.}~\bibnamefont {Kurpas}}, \bibinfo {author} {\bibfnamefont {I.}~\bibnamefont {{\v S}tich}},\ and\ \bibinfo {author} {\bibfnamefont {J.}~\bibnamefont {Fabian}},\ }\bibfield  {title} {\bibinfo {title} {Proximity-enabled control of spin-orbit coupling in phosphorene symmetrically and asymmetrically encapsulated by wse2 monolayers},\ }\href@noop {} {\bibfield  {journal} {\bibinfo  {journal} {Phys. Rev. B}\ }\textbf {\bibinfo {volume} {109}},\ \bibinfo {pages} {075305} (\bibinfo {year} {2024})}\BibitemShut {NoStop}%
\bibitem [{\citenamefont {Milivojevi\ifmmode~\acute{c}\else \'{c}\fi{}}\ \emph {et~al.}(2024)\citenamefont {Milivojevi\ifmmode~\acute{c}\else \'{c}\fi{}}, \citenamefont {Kurpas}, \citenamefont {Rassekh}, \citenamefont {Legut},\ and\ \citenamefont {Gmitra}}]{MKR+24}%
  \BibitemOpen
  \bibfield  {author} {\bibinfo {author} {\bibfnamefont {M.}~\bibnamefont {Milivojevi\ifmmode~\acute{c}\else \'{c}\fi{}}}, \bibinfo {author} {\bibfnamefont {M.}~\bibnamefont {Kurpas}}, \bibinfo {author} {\bibfnamefont {M.}~\bibnamefont {Rassekh}}, \bibinfo {author} {\bibfnamefont {D.}~\bibnamefont {Legut}},\ and\ \bibinfo {author} {\bibfnamefont {M.}~\bibnamefont {Gmitra}},\ }\bibfield  {title} {\bibinfo {title} {Hydrostatic pressure control of the spin-orbit proximity effect, spin relaxation, and thermoelectricity in a phosphorene-${\mathrm{wse}}_{2}$ heterostructure},\ }\href {https://doi.org/10.1103/PhysRevB.110.085306} {\bibfield  {journal} {\bibinfo  {journal} {Phys. Rev. B}\ }\textbf {\bibinfo {volume} {110}},\ \bibinfo {pages} {085306} (\bibinfo {year} {2024})}\BibitemShut {NoStop}%
\bibitem [{\citenamefont {Wei}\ and\ \citenamefont {Peng}(2014)}]{WP14}%
  \BibitemOpen
  \bibfield  {author} {\bibinfo {author} {\bibfnamefont {Q.}~\bibnamefont {Wei}}\ and\ \bibinfo {author} {\bibfnamefont {X.}~\bibnamefont {Peng}},\ }\bibfield  {title} {\bibinfo {title} {{Superior mechanical flexibility of phosphorene and few-layer black phosphorus}},\ }\href {https://doi.org/10.1063/1.4885215} {\bibfield  {journal} {\bibinfo  {journal} {Applied Physics Letters}\ }\textbf {\bibinfo {volume} {104}},\ \bibinfo {pages} {251915} (\bibinfo {year} {2014})},\ \Eprint {https://arxiv.org/abs/https://pubs.aip.org/aip/apl/article-pdf/doi/10.1063/1.4885215/14300079/251915\_1\_online.pdf} {https://pubs.aip.org/aip/apl/article-pdf/doi/10.1063/1.4885215/14300079/251915\_1\_online.pdf} \BibitemShut {NoStop}%
\bibitem [{\citenamefont {Huang}\ \emph {et~al.}(2023)\citenamefont {Huang}, \citenamefont {Faizan}, \citenamefont {Manzoor}, \citenamefont {Brndiar}, \citenamefont {Mitas}, \citenamefont {Fabian},\ and\ \citenamefont {\ifmmode~\check{S}\else \v{S}\fi{}tich}}]{HFM+23}%
  \BibitemOpen
  \bibfield  {author} {\bibinfo {author} {\bibfnamefont {Y.}~\bibnamefont {Huang}}, \bibinfo {author} {\bibfnamefont {A.}~\bibnamefont {Faizan}}, \bibinfo {author} {\bibfnamefont {M.}~\bibnamefont {Manzoor}}, \bibinfo {author} {\bibfnamefont {J.}~\bibnamefont {Brndiar}}, \bibinfo {author} {\bibfnamefont {L.}~\bibnamefont {Mitas}}, \bibinfo {author} {\bibfnamefont {J.}~\bibnamefont {Fabian}},\ and\ \bibinfo {author} {\bibfnamefont {I.}~\bibnamefont {\ifmmode~\check{S}\else \v{S}\fi{}tich}},\ }\bibfield  {title} {\bibinfo {title} {Colossal band gap response of single-layer phosphorene to strain predicted by quantum monte carlo},\ }\href {https://doi.org/10.1103/PhysRevResearch.5.033223} {\bibfield  {journal} {\bibinfo  {journal} {Phys. Rev. Res.}\ }\textbf {\bibinfo {volume} {5}},\ \bibinfo {pages} {033223} (\bibinfo {year} {2023})}\BibitemShut {NoStop}%
\bibitem [{\citenamefont {Li}\ and\ \citenamefont {Appelbaum}(2014)}]{Li2014}%
  \BibitemOpen
  \bibfield  {author} {\bibinfo {author} {\bibfnamefont {P.}~\bibnamefont {Li}}\ and\ \bibinfo {author} {\bibfnamefont {I.}~\bibnamefont {Appelbaum}},\ }\bibfield  {title} {\bibinfo {title} {Electrons and holes in phosphorene},\ }\href {https://doi.org/10.1103/PhysRevB.90.115439} {\bibfield  {journal} {\bibinfo  {journal} {Phys. Rev. B}\ }\textbf {\bibinfo {volume} {90}},\ \bibinfo {pages} {115439} (\bibinfo {year} {2014})}\BibitemShut {NoStop}%
\bibitem [{\citenamefont {Wang}\ \emph {et~al.}(2015{\natexlab{a}})\citenamefont {Wang}, \citenamefont {Kutana}, \citenamefont {Zou},\ and\ \citenamefont {Yakobson}}]{Wang2015}%
  \BibitemOpen
  \bibfield  {author} {\bibinfo {author} {\bibfnamefont {L.}~\bibnamefont {Wang}}, \bibinfo {author} {\bibfnamefont {A.}~\bibnamefont {Kutana}}, \bibinfo {author} {\bibfnamefont {X.}~\bibnamefont {Zou}},\ and\ \bibinfo {author} {\bibfnamefont {B.~I.}\ \bibnamefont {Yakobson}},\ }\bibfield  {title} {\bibinfo {title} {Electro-mechanical anisotropy of phosphorene},\ }\href {https://doi.org/10.1039/C5NR00355E} {\bibfield  {journal} {\bibinfo  {journal} {Nanoscale}\ }\textbf {\bibinfo {volume} {7}},\ \bibinfo {pages} {9746} (\bibinfo {year} {2015}{\natexlab{a}})}\BibitemShut {NoStop}%
\bibitem [{\citenamefont {Luo}\ \emph {et~al.}(2015)\citenamefont {Luo}, \citenamefont {Maassen}, \citenamefont {Deng}, \citenamefont {Du}, \citenamefont {Garrelts}, \citenamefont {Lundstrom}, \citenamefont {Ye},\ and\ \citenamefont {Xu}}]{Luo2015}%
  \BibitemOpen
  \bibfield  {author} {\bibinfo {author} {\bibfnamefont {Z.}~\bibnamefont {Luo}}, \bibinfo {author} {\bibfnamefont {J.}~\bibnamefont {Maassen}}, \bibinfo {author} {\bibfnamefont {Y.}~\bibnamefont {Deng}}, \bibinfo {author} {\bibfnamefont {Y.}~\bibnamefont {Du}}, \bibinfo {author} {\bibfnamefont {R.~P.}\ \bibnamefont {Garrelts}}, \bibinfo {author} {\bibfnamefont {M.~S.}\ \bibnamefont {Lundstrom}}, \bibinfo {author} {\bibfnamefont {P.~D.}\ \bibnamefont {Ye}},\ and\ \bibinfo {author} {\bibfnamefont {X.}~\bibnamefont {Xu}},\ }\bibfield  {title} {\bibinfo {title} {Anisotropic in-plane thermal conductivity observed in few-layer black phosphorus},\ }\href {http://dx.doi.org/10.1038/ncomms9572 http://10.0.4.14/ncomms9572 https://www.nature.com/articles/ncomms9572#supplementary-information} {\bibfield  {journal} {\bibinfo  {journal} {Nature Communications}\ }\textbf {\bibinfo {volume} {6}},\ \bibinfo {pages} {8572} (\bibinfo {year} {2015})}\BibitemShut {NoStop}%
\bibitem [{\citenamefont {Wang}\ \emph {et~al.}(2015{\natexlab{b}})\citenamefont {Wang}, \citenamefont {Jones}, \citenamefont {Seyler}, \citenamefont {Tran}, \citenamefont {Jia}, \citenamefont {Zhao}, \citenamefont {Wang}, \citenamefont {Yang}, \citenamefont {Xu},\ and\ \citenamefont {Xia}}]{Wang2015_exc}%
  \BibitemOpen
  \bibfield  {author} {\bibinfo {author} {\bibfnamefont {X.}~\bibnamefont {Wang}}, \bibinfo {author} {\bibfnamefont {A.~M.}\ \bibnamefont {Jones}}, \bibinfo {author} {\bibfnamefont {K.~L.}\ \bibnamefont {Seyler}}, \bibinfo {author} {\bibfnamefont {V.}~\bibnamefont {Tran}}, \bibinfo {author} {\bibfnamefont {Y.}~\bibnamefont {Jia}}, \bibinfo {author} {\bibfnamefont {H.}~\bibnamefont {Zhao}}, \bibinfo {author} {\bibfnamefont {H.}~\bibnamefont {Wang}}, \bibinfo {author} {\bibfnamefont {L.}~\bibnamefont {Yang}}, \bibinfo {author} {\bibfnamefont {X.}~\bibnamefont {Xu}},\ and\ \bibinfo {author} {\bibfnamefont {F.}~\bibnamefont {Xia}},\ }\bibfield  {title} {\bibinfo {title} {Highly anisotropic and robust excitons in monolayer black phosphorus},\ }\href {https://doi.org/10.1038/nnano.2015.71} {\bibfield  {journal} {\bibinfo  {journal} {Nature Nanotechnology}\ }\textbf {\bibinfo {volume} {10}},\ \bibinfo {pages} {517} (\bibinfo {year} {2015}{\natexlab{b}})}\BibitemShut {NoStop}%
\bibitem [{\citenamefont {Rodin}\ \emph {et~al.}(2014)\citenamefont {Rodin}, \citenamefont {Carvalho},\ and\ \citenamefont {Castro~Neto}}]{Rodin2014}%
  \BibitemOpen
  \bibfield  {author} {\bibinfo {author} {\bibfnamefont {A.~S.}\ \bibnamefont {Rodin}}, \bibinfo {author} {\bibfnamefont {A.}~\bibnamefont {Carvalho}},\ and\ \bibinfo {author} {\bibfnamefont {A.~H.}\ \bibnamefont {Castro~Neto}},\ }\bibfield  {title} {\bibinfo {title} {Strain-induced gap modification in black phosphorus},\ }\href {https://doi.org/10.1103/PhysRevLett.112.176801} {\bibfield  {journal} {\bibinfo  {journal} {Phys. Rev. Lett.}\ }\textbf {\bibinfo {volume} {112}},\ \bibinfo {pages} {176801} (\bibinfo {year} {2014})}\BibitemShut {NoStop}%
\bibitem [{\citenamefont {Elahi}\ \emph {et~al.}(2015)\citenamefont {Elahi}, \citenamefont {Khaliji}, \citenamefont {Tabatabaei}, \citenamefont {Pourfath},\ and\ \citenamefont {Asgari}}]{Elahi2015}%
  \BibitemOpen
  \bibfield  {author} {\bibinfo {author} {\bibfnamefont {M.}~\bibnamefont {Elahi}}, \bibinfo {author} {\bibfnamefont {K.}~\bibnamefont {Khaliji}}, \bibinfo {author} {\bibfnamefont {S.~M.}\ \bibnamefont {Tabatabaei}}, \bibinfo {author} {\bibfnamefont {M.}~\bibnamefont {Pourfath}},\ and\ \bibinfo {author} {\bibfnamefont {R.}~\bibnamefont {Asgari}},\ }\bibfield  {title} {\bibinfo {title} {Modulation of electronic and mechanical properties of phosphorene through strain},\ }\href {https://doi.org/10.1103/PhysRevB.91.115412} {\bibfield  {journal} {\bibinfo  {journal} {Phys. Rev. B}\ }\textbf {\bibinfo {volume} {91}},\ \bibinfo {pages} {115412} (\bibinfo {year} {2015})}\BibitemShut {NoStop}%
\bibitem [{\citenamefont {Ge}\ \emph {et~al.}(2015)\citenamefont {Ge}, \citenamefont {Wan}, \citenamefont {Yang},\ and\ \citenamefont {Yao}}]{Ge_2015}%
  \BibitemOpen
  \bibfield  {author} {\bibinfo {author} {\bibfnamefont {Y.}~\bibnamefont {Ge}}, \bibinfo {author} {\bibfnamefont {W.}~\bibnamefont {Wan}}, \bibinfo {author} {\bibfnamefont {F.}~\bibnamefont {Yang}},\ and\ \bibinfo {author} {\bibfnamefont {Y.}~\bibnamefont {Yao}},\ }\bibfield  {title} {\bibinfo {title} {The strain effect on superconductivity in phosphorene: a first-principles prediction},\ }\href {https://doi.org/10.1088/1367-2630/17/3/035008} {\bibfield  {journal} {\bibinfo  {journal} {New Journal of Physics}\ }\textbf {\bibinfo {volume} {17}},\ \bibinfo {pages} {035008} (\bibinfo {year} {2015})}\BibitemShut {NoStop}%
\bibitem [{\citenamefont {Giannozzi}\ \emph {et~al.}(2009)\citenamefont {Giannozzi}, \citenamefont {Baroni}, \citenamefont {Bonini}, \citenamefont {Calandra}, \citenamefont {Car}, \citenamefont {Cavazzoni}, \citenamefont {Ceresoli}, \citenamefont {Chiarotti}, \citenamefont {Cococcioni}, \citenamefont {Dabo} \emph {et~al.}}]{QE-2009}%
  \BibitemOpen
  \bibfield  {author} {\bibinfo {author} {\bibfnamefont {P.}~\bibnamefont {Giannozzi}}, \bibinfo {author} {\bibfnamefont {S.}~\bibnamefont {Baroni}}, \bibinfo {author} {\bibfnamefont {N.}~\bibnamefont {Bonini}}, \bibinfo {author} {\bibfnamefont {M.}~\bibnamefont {Calandra}}, \bibinfo {author} {\bibfnamefont {R.}~\bibnamefont {Car}}, \bibinfo {author} {\bibfnamefont {C.}~\bibnamefont {Cavazzoni}}, \bibinfo {author} {\bibfnamefont {D.}~\bibnamefont {Ceresoli}}, \bibinfo {author} {\bibfnamefont {G.~L.}\ \bibnamefont {Chiarotti}}, \bibinfo {author} {\bibfnamefont {M.}~\bibnamefont {Cococcioni}}, \bibinfo {author} {\bibfnamefont {I.}~\bibnamefont {Dabo}}, \emph {et~al.},\ }\bibfield  {title} {\bibinfo {title} {{QUANTUM ESPRESSO: a modular and open-source software project for quantum simulations of materials}},\ }\href {http://www.quantum-espresso.org} {\bibfield  {journal} {\bibinfo  {journal} {J. Phys.: Condens. Matter}\ }\textbf {\bibinfo {volume} {21}},\ \bibinfo {pages} {395502} (\bibinfo {year}
  {2009})}\BibitemShut {NoStop}%
\bibitem [{\citenamefont {Giannozzi}\ \emph {et~al.}(2017)\citenamefont {Giannozzi}, \citenamefont {Andreussi}, \citenamefont {Brumme}, \citenamefont {Bunau}, \citenamefont {Nardelli}, \citenamefont {Calandra}, \citenamefont {Car}, \citenamefont {Cavazzoni}, \citenamefont {Ceresoli}, \citenamefont {Cococcioni} \emph {et~al.}}]{Giannozzi_2017}%
  \BibitemOpen
  \bibfield  {author} {\bibinfo {author} {\bibfnamefont {P.}~\bibnamefont {Giannozzi}}, \bibinfo {author} {\bibfnamefont {O.}~\bibnamefont {Andreussi}}, \bibinfo {author} {\bibfnamefont {T.}~\bibnamefont {Brumme}}, \bibinfo {author} {\bibfnamefont {O.}~\bibnamefont {Bunau}}, \bibinfo {author} {\bibfnamefont {M.~B.}\ \bibnamefont {Nardelli}}, \bibinfo {author} {\bibfnamefont {M.}~\bibnamefont {Calandra}}, \bibinfo {author} {\bibfnamefont {R.}~\bibnamefont {Car}}, \bibinfo {author} {\bibfnamefont {C.}~\bibnamefont {Cavazzoni}}, \bibinfo {author} {\bibfnamefont {D.}~\bibnamefont {Ceresoli}}, \bibinfo {author} {\bibfnamefont {M.}~\bibnamefont {Cococcioni}}, \emph {et~al.},\ }\bibfield  {title} {\bibinfo {title} {{Advanced capabilities for materials modelling with Quantum ESPRESSO}},\ }\href {https://doi.org/10.1088/1361-648X/aa8f79} {\bibfield  {journal} {\bibinfo  {journal} {Journal of Physics: Condensed Matter}\ }\textbf {\bibinfo {volume} {29}},\ \bibinfo {pages} {465901} (\bibinfo {year} {2017})}\BibitemShut
  {NoStop}%
\bibitem [{\citenamefont {Perdew}\ \emph {et~al.}(1996)\citenamefont {Perdew}, \citenamefont {Burke},\ and\ \citenamefont {Ernzerhof}}]{perdew1996}%
  \BibitemOpen
  \bibfield  {author} {\bibinfo {author} {\bibfnamefont {J.~P.}\ \bibnamefont {Perdew}}, \bibinfo {author} {\bibfnamefont {K.}~\bibnamefont {Burke}},\ and\ \bibinfo {author} {\bibfnamefont {M.}~\bibnamefont {Ernzerhof}},\ }\href {https://doi.org/10.1103/PhysRevLett.100.136406} {\bibfield  {journal} {\bibinfo  {journal} {Phys. Rev. Lett.}\ }\textbf {\bibinfo {volume} {77}},\ \bibinfo {pages} {3865} (\bibinfo {year} {1996})}\BibitemShut {NoStop}%
\bibitem [{\citenamefont {Hamann}(2013)}]{Hamman2017}%
  \BibitemOpen
  \bibfield  {author} {\bibinfo {author} {\bibfnamefont {D.~R.}\ \bibnamefont {Hamann}},\ }\bibfield  {title} {\bibinfo {title} {Optimized norm-conserving vanderbilt pseudopotentials},\ }\href {https://doi.org/10.1103/PhysRevB.88.085117} {\bibfield  {journal} {\bibinfo  {journal} {Phys. Rev. B}\ }\textbf {\bibinfo {volume} {88}},\ \bibinfo {pages} {085117} (\bibinfo {year} {2013})}\BibitemShut {NoStop}%
\bibitem [{\citenamefont {Scherpelz}\ \emph {et~al.}(2016)\citenamefont {Scherpelz}, \citenamefont {Govoni}, \citenamefont {Hamada},\ and\ \citenamefont {Galli}}]{Scherpelz2016}%
  \BibitemOpen
  \bibfield  {author} {\bibinfo {author} {\bibfnamefont {P.}~\bibnamefont {Scherpelz}}, \bibinfo {author} {\bibfnamefont {M.}~\bibnamefont {Govoni}}, \bibinfo {author} {\bibfnamefont {I.}~\bibnamefont {Hamada}},\ and\ \bibinfo {author} {\bibfnamefont {G.}~\bibnamefont {Galli}},\ }\bibfield  {title} {\bibinfo {title} {{Implementation and Validation of Fully Relativistic GW Calculations: Spin–Orbit Coupling in Molecules, Nanocrystals, and Solids}},\ }\href {https://doi.org/10.1021/acs.jctc.6b00114} {\bibfield  {journal} {\bibinfo  {journal} {J. Chem. Theory Comput.}\ }\textbf {\bibinfo {volume} {12}},\ \bibinfo {pages} {3523} (\bibinfo {year} {2016})}\BibitemShut {NoStop}%
\bibitem [{\citenamefont {Bengtsson}(1999)}]{B99}%
  \BibitemOpen
  \bibfield  {author} {\bibinfo {author} {\bibfnamefont {L.}~\bibnamefont {Bengtsson}},\ }\bibfield  {title} {\bibinfo {title} {Dipole correction for surface supercell calculations},\ }\href {https://doi.org/10.1103/PhysRevB.59.12301} {\bibfield  {journal} {\bibinfo  {journal} {Phys. Rev. B}\ }\textbf {\bibinfo {volume} {59}},\ \bibinfo {pages} {12301} (\bibinfo {year} {1999})}\BibitemShut {NoStop}%
\bibitem [{\citenamefont {Monkhorst}\ and\ \citenamefont {Pack}(1976)}]{MPack}%
  \BibitemOpen
  \bibfield  {author} {\bibinfo {author} {\bibfnamefont {H.~J.}\ \bibnamefont {Monkhorst}}\ and\ \bibinfo {author} {\bibfnamefont {J.~D.}\ \bibnamefont {Pack}},\ }\bibfield  {title} {\bibinfo {title} {Special points for brillouin-zone integrations},\ }\href {https://doi.org/10.1103/PhysRevB.13.5188} {\bibfield  {journal} {\bibinfo  {journal} {Phys. Rev. B}\ }\textbf {\bibinfo {volume} {13}},\ \bibinfo {pages} {5188} (\bibinfo {year} {1976})}\BibitemShut {NoStop}%
\bibitem [{\citenamefont {Elliott}(1954)}]{EY1954}%
  \BibitemOpen
  \bibfield  {author} {\bibinfo {author} {\bibfnamefont {R.~J.}\ \bibnamefont {Elliott}},\ }\bibfield  {title} {\bibinfo {title} {Theory of the {Effect} of {Spin}-{Orbit} {Coupling} on {Magnetic} {Resonance} in {Some} {Semiconductors}},\ }\href {https://doi.org/10.1103/PhysRev.96.266} {\bibfield  {journal} {\bibinfo  {journal} {Phys. Rev.}\ }\textbf {\bibinfo {volume} {96}},\ \bibinfo {pages} {266} (\bibinfo {year} {1954})}\BibitemShut {NoStop}%
\bibitem [{\citenamefont {Zimmermann}\ \emph {et~al.}(2012)\citenamefont {Zimmermann}, \citenamefont {Mavropoulos}, \citenamefont {Heers}, \citenamefont {Long}, \citenamefont {Bl{\" u}gel},\ and\ \citenamefont {Mokrousov}}]{Zimmermann}%
  \BibitemOpen
  \bibfield  {author} {\bibinfo {author} {\bibfnamefont {B.}~\bibnamefont {Zimmermann}}, \bibinfo {author} {\bibfnamefont {P.}~\bibnamefont {Mavropoulos}}, \bibinfo {author} {\bibfnamefont {S.}~\bibnamefont {Heers}}, \bibinfo {author} {\bibfnamefont {N.~H.}\ \bibnamefont {Long}}, \bibinfo {author} {\bibfnamefont {S.}~\bibnamefont {Bl{\" u}gel}},\ and\ \bibinfo {author} {\bibfnamefont {Y.}~\bibnamefont {Mokrousov}},\ }\bibfield  {title} {\bibinfo {title} {Anisotropy of spin relaxation in metals},\ }\href {https://doi.org/10.1103/PhysRevLett.109.236603} {\bibfield  {journal} {\bibinfo  {journal} {Physical Review Letters}\ }\textbf {\bibinfo {volume} {109}},\ \bibinfo {pages} {236603} (\bibinfo {year} {2012})}\BibitemShut {NoStop}%
\bibitem [{\citenamefont {Kurpas}\ \emph {et~al.}(2019)\citenamefont {Kurpas}, \citenamefont {Faria~Junior}, \citenamefont {Gmitra},\ and\ \citenamefont {Fabian}}]{KFGF19}%
  \BibitemOpen
  \bibfield  {author} {\bibinfo {author} {\bibfnamefont {M.}~\bibnamefont {Kurpas}}, \bibinfo {author} {\bibfnamefont {P.~E.}\ \bibnamefont {Faria~Junior}}, \bibinfo {author} {\bibfnamefont {M.}~\bibnamefont {Gmitra}},\ and\ \bibinfo {author} {\bibfnamefont {J.}~\bibnamefont {Fabian}},\ }\bibfield  {title} {\bibinfo {title} {Spin-orbit coupling in elemental two-dimensional materials},\ }\href {https://doi.org/10.1103/PhysRevB.100.125422} {\bibfield  {journal} {\bibinfo  {journal} {Phys. Rev. B}\ }\textbf {\bibinfo {volume} {100}},\ \bibinfo {pages} {125422} (\bibinfo {year} {2019})}\BibitemShut {NoStop}%
\bibitem [{\citenamefont {Fabian}\ and\ \citenamefont {Das~Sarma}(1998)}]{fabian_spin_1998}%
  \BibitemOpen
  \bibfield  {author} {\bibinfo {author} {\bibfnamefont {J.}~\bibnamefont {Fabian}}\ and\ \bibinfo {author} {\bibfnamefont {S.}~\bibnamefont {Das~Sarma}},\ }\bibfield  {title} {\bibinfo {title} {Spin {Relaxation} of {Conduction} {Electrons} in {Polyvalent} {Metals}: {Theory} and a {Realistic} {Calculation}},\ }\href {https://doi.org/10.1103/PhysRevLett.81.5624} {\bibfield  {journal} {\bibinfo  {journal} {Phys. Rev. Lett.}\ }\textbf {\bibinfo {volume} {81}},\ \bibinfo {pages} {5624} (\bibinfo {year} {1998})}\BibitemShut {NoStop}%
\bibitem [{\citenamefont {Heyd}\ \emph {et~al.}(2003)\citenamefont {Heyd}, \citenamefont {Scuseria},\ and\ \citenamefont {Ernzerhof}}]{HSE}%
  \BibitemOpen
  \bibfield  {author} {\bibinfo {author} {\bibfnamefont {J.}~\bibnamefont {Heyd}}, \bibinfo {author} {\bibfnamefont {G.~E.}\ \bibnamefont {Scuseria}},\ and\ \bibinfo {author} {\bibfnamefont {M.}~\bibnamefont {Ernzerhof}},\ }\bibfield  {title} {\bibinfo {title} {{Hybrid functionals based on a screened Coulomb potential}},\ }\href {https://doi.org/10.1063/1.1564060} {\bibfield  {journal} {\bibinfo  {journal} {The Journal of Chemical Physics}\ }\textbf {\bibinfo {volume} {118}},\ \bibinfo {pages} {8207} (\bibinfo {year} {2003})}\BibitemShut {NoStop}%
\bibitem [{\citenamefont {Popović}\ \emph {et~al.}(2015)\citenamefont {Popović}, \citenamefont {Kurdestany},\ and\ \citenamefont {Satpathy}}]{PKS15}%
  \BibitemOpen
  \bibfield  {author} {\bibinfo {author} {\bibfnamefont {Z.~S.}\ \bibnamefont {Popović}}, \bibinfo {author} {\bibfnamefont {J.~M.}\ \bibnamefont {Kurdestany}},\ and\ \bibinfo {author} {\bibfnamefont {S.}~\bibnamefont {Satpathy}},\ }\bibfield  {title} {\bibinfo {title} {{Electronic structure and anisotropic Rashba spin-orbit coupling in monolayer black phosphorus}},\ }\href {https://doi.org/10.1103/PhysRevB.92.0351356} {\bibfield  {journal} {\bibinfo  {journal} {Physical Review B}\ }\textbf {\bibinfo {volume} {92}},\ \bibinfo {pages} {035135} (\bibinfo {year} {2015})}\BibitemShut {NoStop}%
\bibitem [{\citenamefont {{Dyakonov}}\ and\ \citenamefont {{Perel}}(1971)}]{dyakonov_1971R}%
  \BibitemOpen
  \bibfield  {author} {\bibinfo {author} {\bibfnamefont {M.~I.}\ \bibnamefont {{Dyakonov}}}\ and\ \bibinfo {author} {\bibfnamefont {V.~I.}\ \bibnamefont {{Perel}}},\ }\bibfield  {title} {\bibinfo {title} {{Spin relaxation of conduction electrons in noncentrosymmetric semiconductors}},\ }\href@noop {} {\bibfield  {journal} {\bibinfo  {journal} {Sov. Phys. Solid. State}\ }\textbf {\bibinfo {volume} {13}},\ \bibinfo {pages} {3023} (\bibinfo {year} {1971})}\BibitemShut {NoStop}%
\end{thebibliography}%

\end{document}


\title{Supplemental Material: Strain-tuning of spin anisotropy in single-layer phosphorene: insights from Elliott-Yafet and Dyakonov-Perel spin relaxation rates}

\author{Paulina Jureczko}
\email{paulina.jureczko@us.edu.pl}
\affiliation{Institute of Physics, University of Silesia in Katowice, 41-500 Chorzów, Poland}%
\author{Marko Milivojevi\' c}
\affiliation{Institute of Informatics, Slovak Academy of Sciences, 84507 Bratislava, Slovakia}
\affiliation {Faculty of Physics, University of Belgrade, 11001 Belgrade, Serbia}
\author{Marcin Kurpas}
\email{marcin.kurpas@us.edu.pl}
\affiliation{Institute of Physics, University of Silesia in Katowice, 41-500 Chorzów, Poland}%

\maketitle

\section{Effect of the strain on crystal structure}

Here we briefly discuss the effects of the strain applied in the zigzag and armchair of the phosphorene monolayer on phosphorene's basis physical properties. 

\begin{figure*}[h]
    \centering
    \includegraphics[width=0.75\columnwidth]{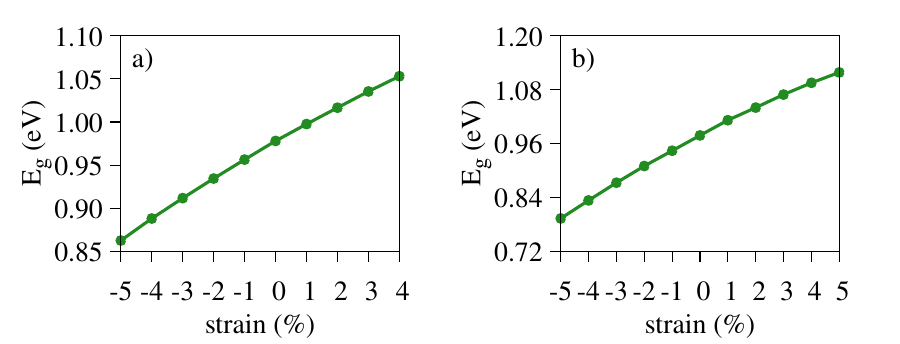}
    \caption{Band gap size dependency from tensile and compressive strain along zigzag~a) and armchair~b) directions. For all strains in each direction the conduction band minimum remains at $\Gamma$ point. Maximum of valence band is moved toward X point after inclusion of the compressive strain to the system, while for tensile strain the maximum stays at $\Gamma$.  }
    \label{figsupp:gap_x_y}
\end{figure*}

In FIG.~\ref{figsupp:gap_x_y}, we present the dependency of the band gap on the applied strain in the zigzag~a) and armchair~b) direction. In both cases, the plots show a linear dependence of the band gap on the applied strain,
where the conduction band minima are located at the $\Gamma$ point, whereas the position of the valence band maxima is strain dependent; for compressive strain, it moves towards the $X$ point, while for tensile strain it stays at the $\Gamma$ point.

\begin{figure*}[h]
    \centering
    \includegraphics[width=0.75\columnwidth]{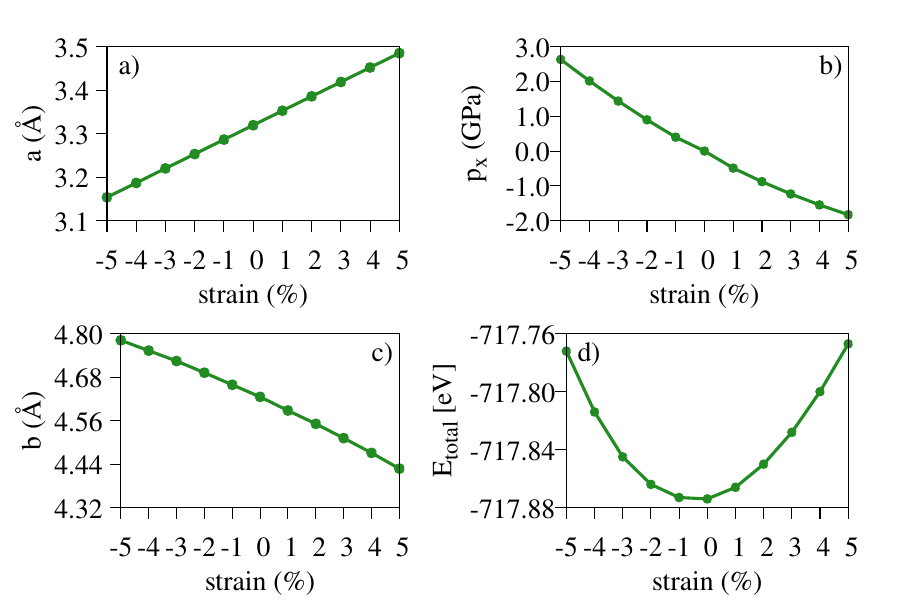}
    \caption{
    In the case of the phosphorene monolayer strained in the zigzag (x) direction, characterized by the intensity change of the parameter {\bf a}~a), the induced pressure $p_x$ in the zigzag direction~b), intensity change of the lattice parameter in the armchair direction~c), and the total energy change~d) is given.}
    \label{figsupp:paramx}
\end{figure*}

In FIG.~\ref{figsupp:paramx}, after straining the phosphorene in the zigzag direction~a), we show that the induced pressure in the zigzag direction~b), change of the lattice parameter in the armchair direction~c) follow the linear dependence on the applied strain. Moreover, in~d), we present the dependence of the total energy on the applied strain; it is clear that the zero strain case scenario is the one with the minimal energy. 

\begin{figure*}[h]
    \centering
    \includegraphics[width=0.75\columnwidth]{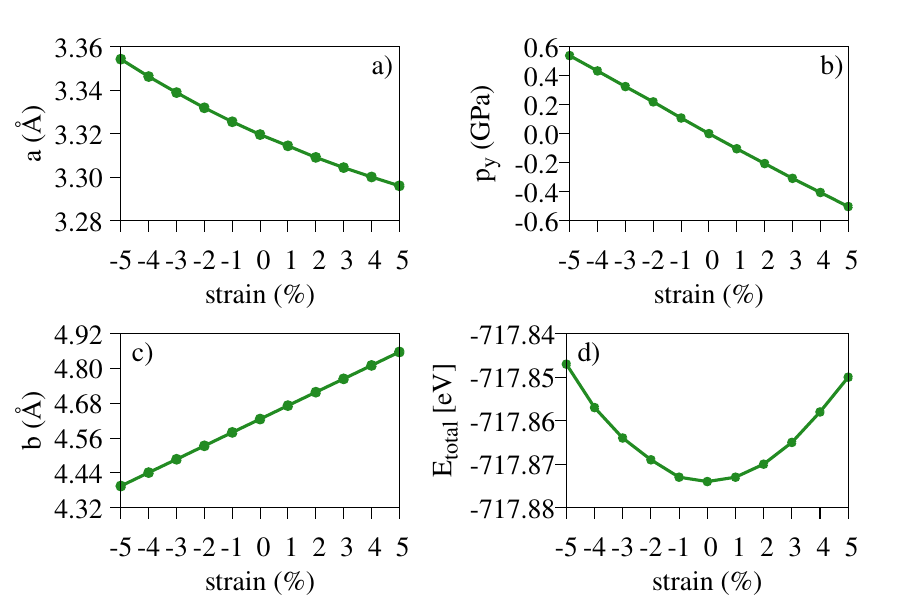}
    \caption{
    In the case of the phosphorene monolayer strained in the armchair (y) direction, characterized by the intensity change of the parameter {\bf b}~a), the induced pressure $p_y$ in the armchair direction~b), intensity change of the lattice parameter in the zigzag direction~c), and the total energy change~d) is given. 
    }
    \label{figsupp:paramy}
\end{figure*}

Finally, in FIG.~\ref{figsupp:paramy} we represent the same response function as in the previous Figure, assuming that the applied strain is now in the armchair direction. The result show that both the pressure~b) in the armchair direction and the induced pressure in the zigzag direction~c), follow the linear dependence on the applied strain. Additionally, the total energy change~d) due to the applied strain follows the quadratic dependence, with the minima located at the zero strain case, as expected.

\section{Strain effects on the spin-mixing parameter $b^2$}

In this section we investigate the effects of external strain on intrinsic SOC in phosphorene. This will be done by studying the Elliott spin mixing parameter $b^2$ which is able to provide the description of SOC in centrosymmetric non-magnetic materials. 

\begin{figure*}[t]
    \centering
    \includegraphics[width=0.9\columnwidth]{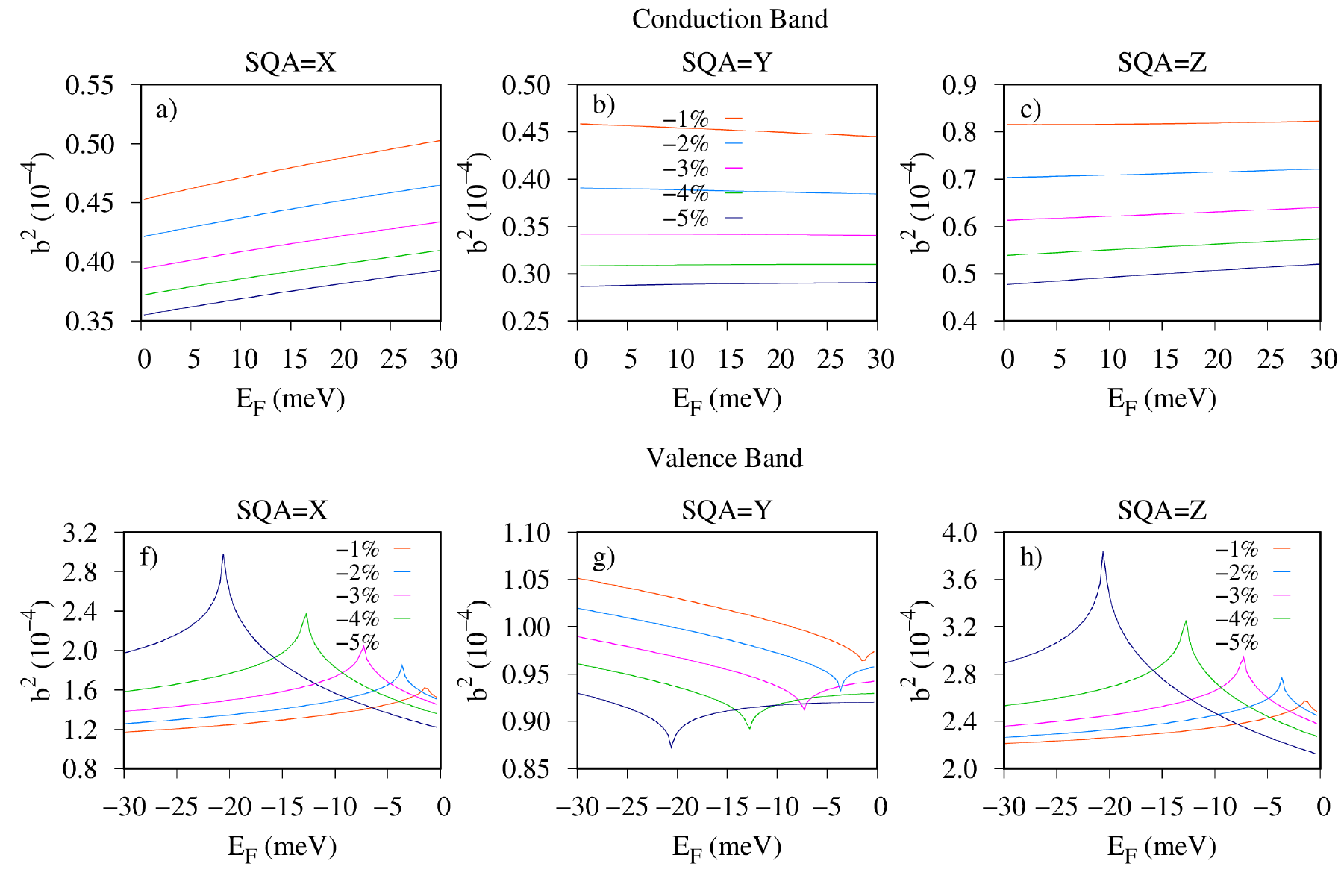}
    \caption{Average values of spin-mixing parameter $b^2$ of phosphorene electrons~a)-c) and holes~d)-f) for each compressive strain applied along zigzag\,(x) direction.}
    \label{figsupp:b2_comp_zigzag}
\end{figure*}

\begin{figure*}[h]
    \centering
    \includegraphics[width=0.9\columnwidth]{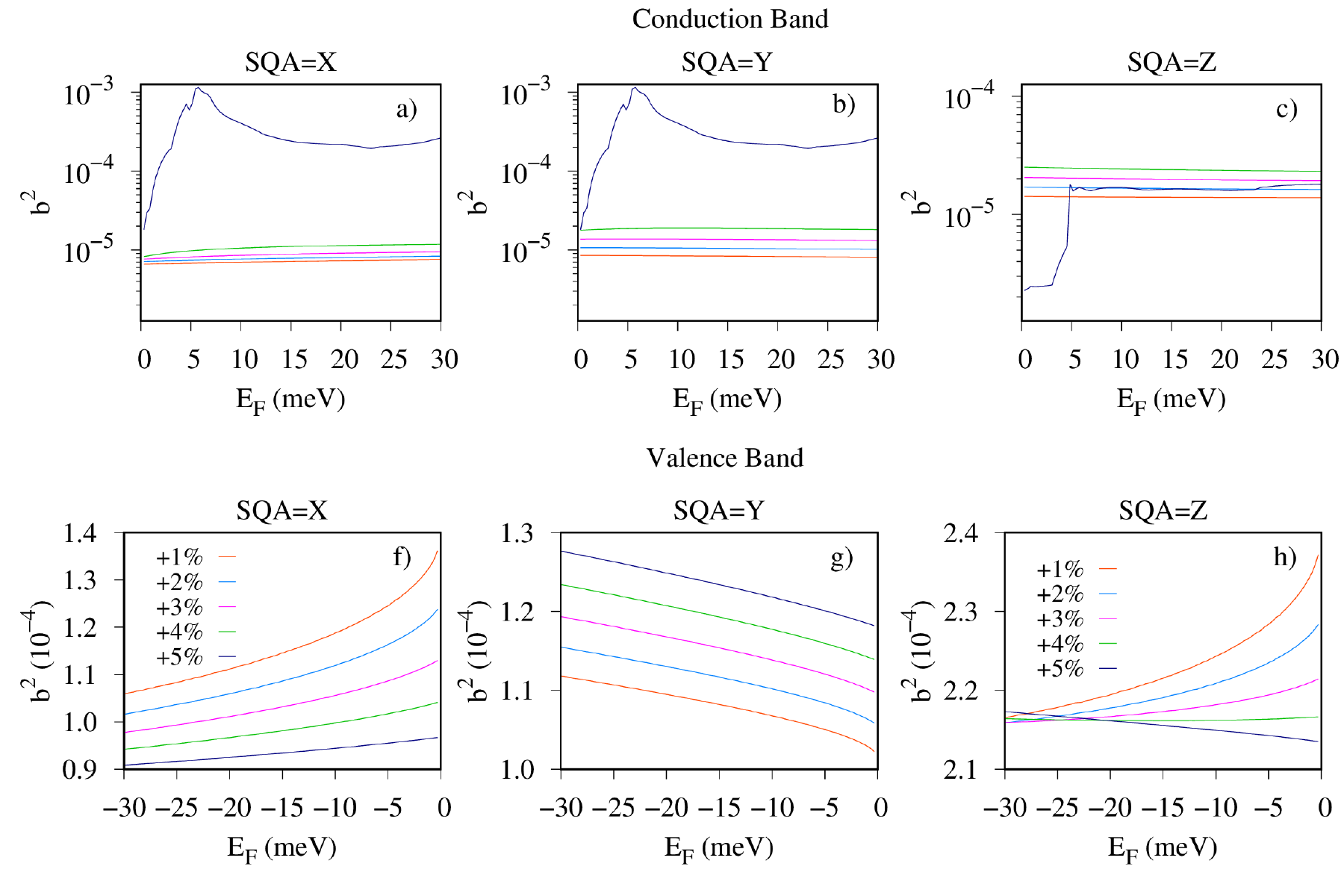}
    \caption{Average values of spin-mixing parameter $b^2$ of phosphorene electrons~a)-c) and holes~d)-f) for each tensile strain applied along zigzag\,(x) direction.}
    \label{figsupp:b2_tens_zigzag}
\end{figure*}

In Fig.~\ref{figsupp:b2_comp_zigzag} we show the average spin mixing parameter $b^2$ in the conduction~a)-c) and valence~d)-f) band plotted versus the Fermi energy for compressive strain along the zigzag (x) direction. Whereas in the case of conductance band the changes due to strain are minimal, in the case of the valence band are more pronounced. Moreover, for the same strain, the position of the peak for each of the component of $b^2$ appears at the same energy, suggesting that its origin comes from the position of the valence band maxima. The change of the position of the peak due to the applied strain is a consequence of the change in the position of the valence band maxima with the applied strain.

Furthermore, in FIG.~\ref{figsupp:b2_tens_zigzag} we show the average spin mixing parameter $b^2$ in the conduction~a)-c) and valence~d)-f) band plotted versus the Fermi energy for the tensile strain along the zigzag direction. Whereas the cases Fig.~\ref{figsupp:b2_tens_zigzag}~a)-c) have been discussed in the main text of the manuscript, we mention that there are no significant changes of $b^2$ of phosphorene holes due to the applied positive strain along the zigzag direction.

\begin{figure*}[h]
    \centering
    \includegraphics[width=0.9\columnwidth]{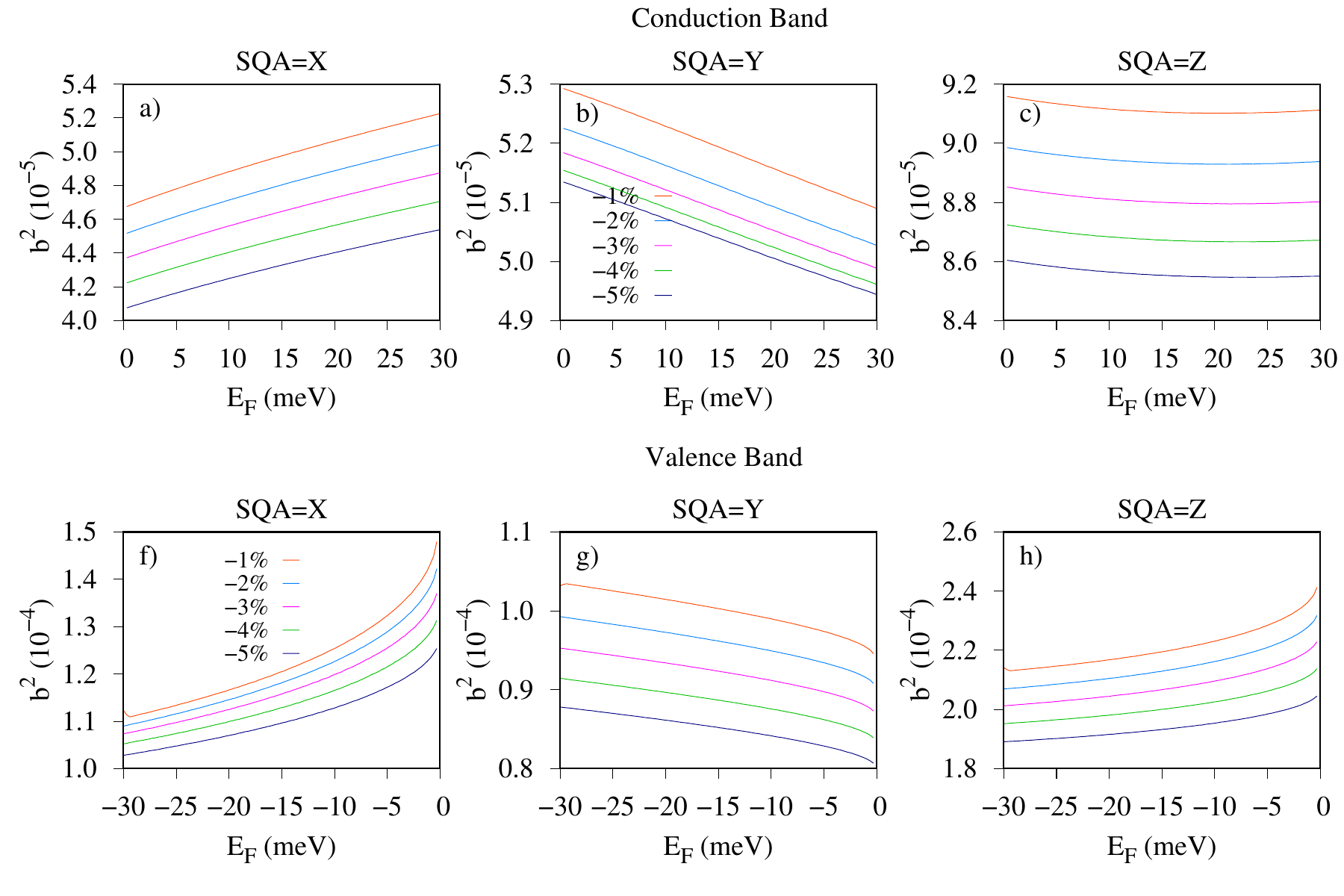}
    \caption{Average values of spin-mixing parameter $b^2$ of phosphorene electrons~a)-c) and holes~d)-f) for each compressive strain applied along armchair\,(y) direction.}
    \label{figsupp:b2_comp_amrchair}
\end{figure*}

\begin{figure*}[h]
    \centering
    \includegraphics[width=0.9\columnwidth]{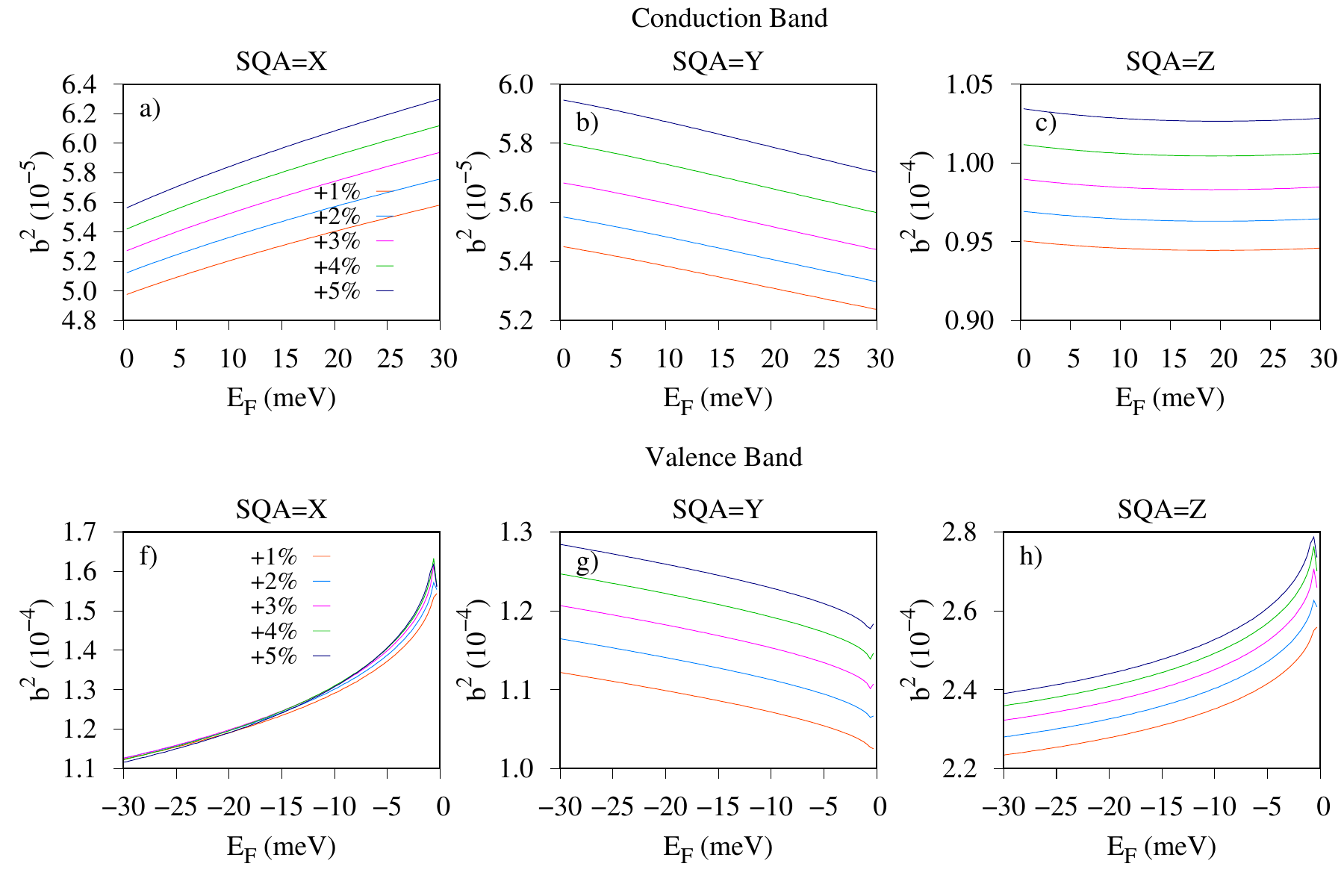}
    \caption{Average values of spin-mixing parameter $b^2$ of phosphorene electrons~a)-c) and holes~d)-f) for each tensile strain applied along armchair\,(y) direction.}
    \label{figsupp:b2_tens_amrchair}
\end{figure*}

The dependence of the spin mixing parameter $b^2$ in the conductance and valence band on the compressive and tensile strain in the armchair (y) direction is given in FIG.~\ref{figsupp:b2_comp_amrchair} and FIG.~\ref{figsupp:b2_tens_amrchair}, respectively.
In all cases, the strain has a minimal effect on the spin mixing parameter. 

\section{Rashba effect in phosphorene strained in the armchair direction}
\begin{figure*}[t]
    \centering
    \includegraphics[angle=270, width=0.99\textwidth]{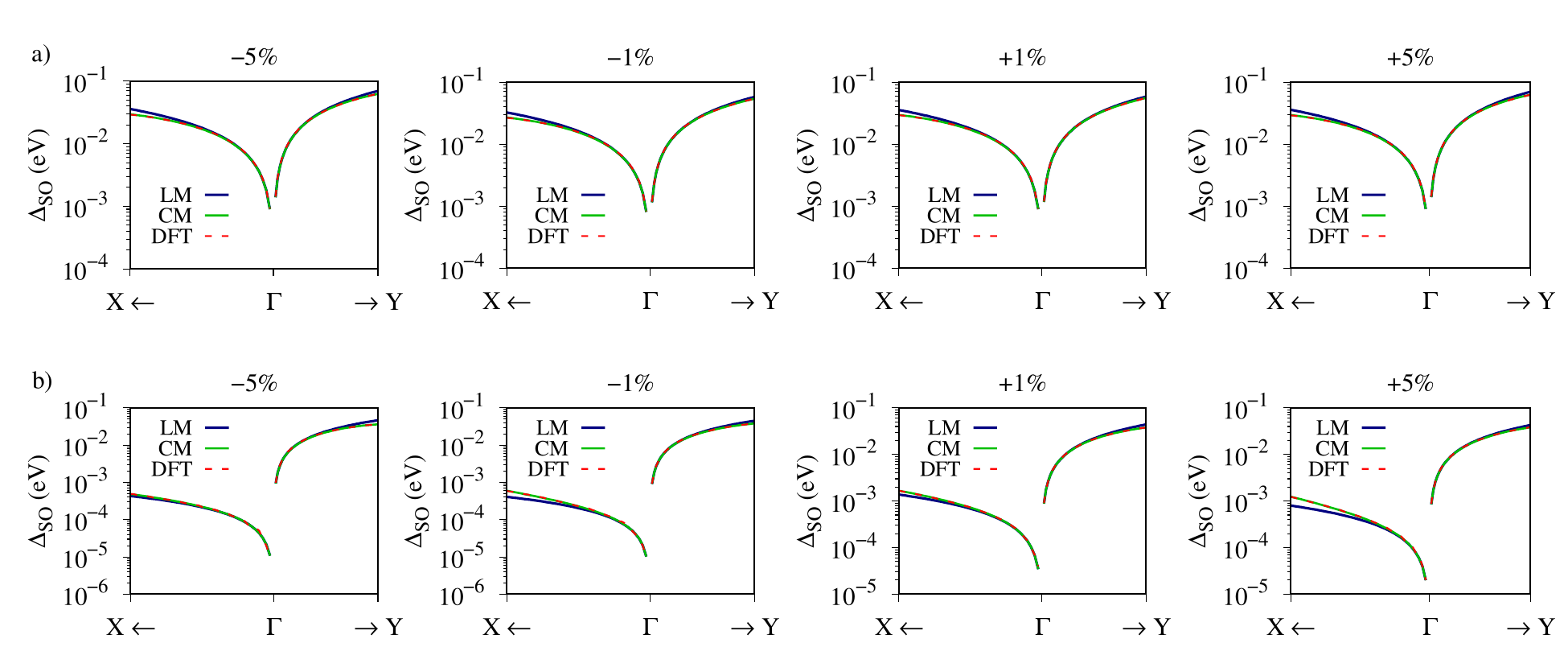}
      \caption{Spin splittings of the top valence~a) and the bottom conductance~b) band along the X$\Gamma$Y path are analyzed for different strain values and the electric field E=1V/nm. 
      In all cases, we compare
      the effective spin-orbit Hamiltonian model in the linear ($\propto k$) and cubic approximation (terms $\propto k$ and $\propto k^3$ are included)
      and the DFT data,
      for $k$-values up to the $5\%$ of both the $\Gamma$X and $\Gamma$Y lines. The comparison show that the cubic approximation is valid for the analyzed $k$-range and that the linear approximation captures the dominant spin physics.}
    \label{fig:LINCUBDFT}
\end{figure*}

In this section we analyze the effects of the applied electric field on the spin physics of phosphorene strained in the armchair (y) direction. The analyzed system has the ${\bf C}_{2{\rm v}}$ symmetry and based on such a symmetry one can derive the spin-orbit Hamiltonian model that successfully describes the spin physics of the top valence band and the bottom conductance band around the $\Gamma$ point
\begin{eqnarray}\label{effmodel}
    H_{\rm SOC}^{\rm eff}&=& \lambda_1 k_x \sigma_y + \lambda_2 k_y \sigma_x+\lambda_3 k_x^3 \sigma_y+
    \lambda_4 k_y^3 \sigma_x+ \lambda_5 k_xk_y^2 \sigma_y
    + \lambda_6 k_x^2k_y \sigma_x,
\end{eqnarray}
up to the terms qubic in momenta (for more details, see the first paragraph of Section IV in the main text). Since the terms linear in momenta represent the dominant contribution to the spin splitting up the 5$\%$ of the $\Gamma$X and the $\Gamma$Y line, see FIG.~\ref{fig:LINCUBDFT}, in what follows we will focus on the contribution of the terms proportional to $\lambda_1$ and $\lambda_2$ parameters. 

\begin{figure*}[t]
    \centering
    \includegraphics[angle=270, width=0.99\textwidth]{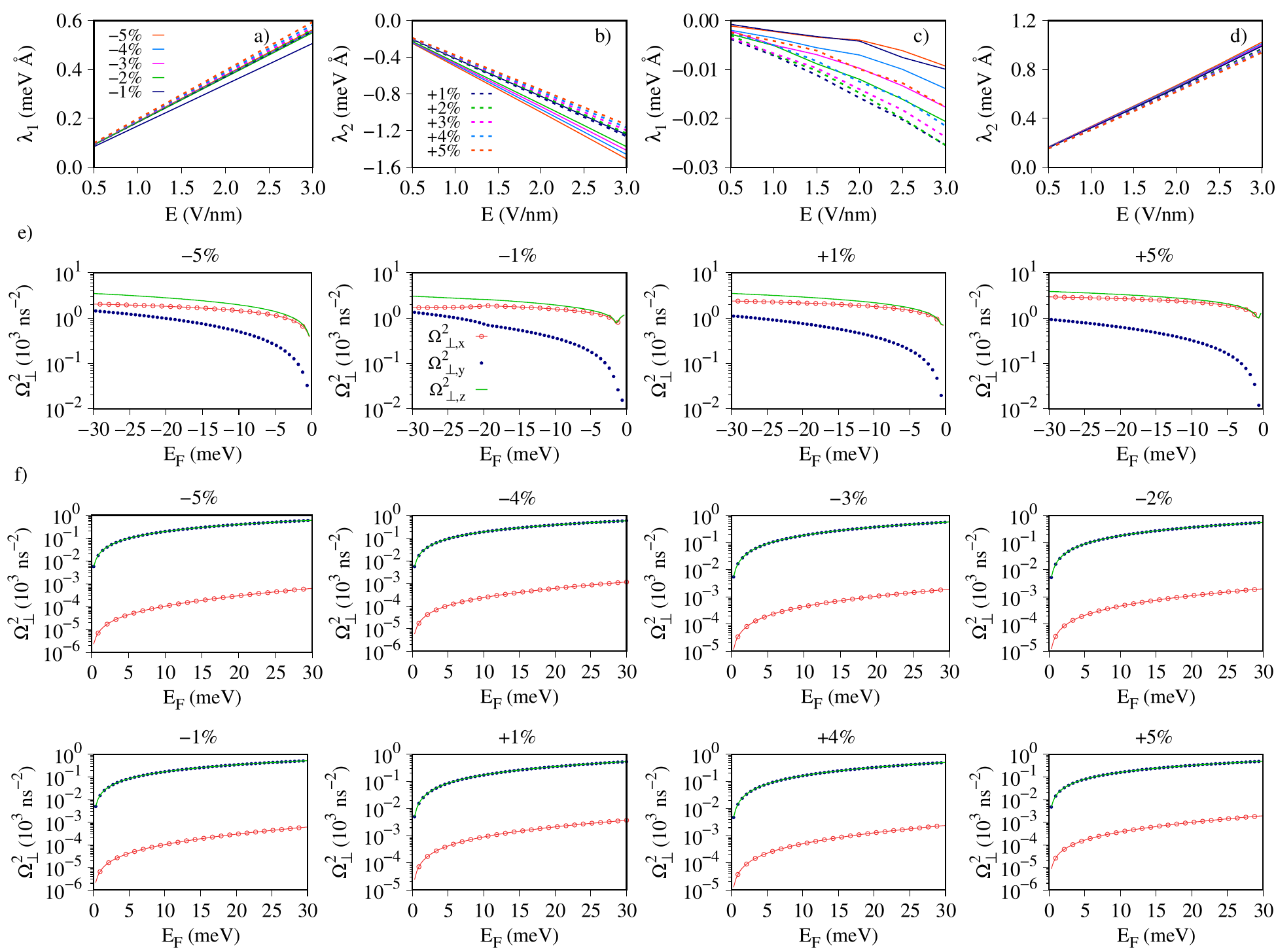}
      \caption{{\it Influence of electric field on spin physics of phosphorene strained in the armchair direction}. Parameters of linear terms $\lambda_{1}$, $\lambda_{2}$ for valence~a), b) and conduction band~c), d) are obtained after fitting the DFT data to the Hamiltonian model given in the main text. Additionally, for the applied electric field of E=1V/nm and different values of strain, we present the dependence of squared spin-orbit field components $\Omega_{\perp,i}^2$, $i = \{x, y, z\}$, for phosphorene hole e) and electron f) states, against the position of the Fermi level $E_{\rm F}$ measured from the top/bottom of the valence/conduction band.}
    \label{fig:Rashba}
\end{figure*}

In FIG.~\ref{fig:Rashba} we present the dependence of the effective spin parameters $\lambda_1$ and $\lambda_2$ on the electric field strength of phosphorene holes, FIG.~\ref{fig:Rashba}~a)-b), and electrons, FIG.~\ref{fig:Rashba}~c)-d). Our results indicate that the strain effects are weak in the case of phosphorene holes and the term $\propto\lambda_2 k_y \sigma_x$ in the case of phosphorene electrons; strain effects of the phosphorene electron's term $\propto\lambda_1 k_x \sigma_y$ are more pronounced.

In FIG.~\ref{fig:Rashba}~e) and f) the dependence of the  squared spin-orbit field components $\Omega_{\perp,i}^2$, $i = \{x, y, z\}$, against the position of the Fermi level $E_{\rm F}$ measured from the top/bottom of the valence/conduction band is given. In all cases, the electric field strength of 1 V/nm is assumed. For phosphorene holes, FIG.~\ref{fig:Rashba}~e), we see a negligible change of spin-orbit field components, as expected due to the minimal response of spin-orbit parameters to both the compressive and tensile strain. Whereas the same happens to the squared spin-orbit field components $\Omega_{\perp,y/z}^2$ of phosphorene electrons due to the minimal changes of the term $\propto\lambda_2 k_y\sigma_x$ by varying strain, in the case of the spin-orbit field component the variation of the $\Omega_{\perp,x}^2$ is more pronounced and can be connected to the more sizable changes of the  $\propto\lambda_1 k_x\sigma_y$ term with strain modification.

\section{Comparison between different exchange-correlation functionals}

\begin{figure*}[t]
    \centering
    \includegraphics[angle=270, width=0.99\textwidth]{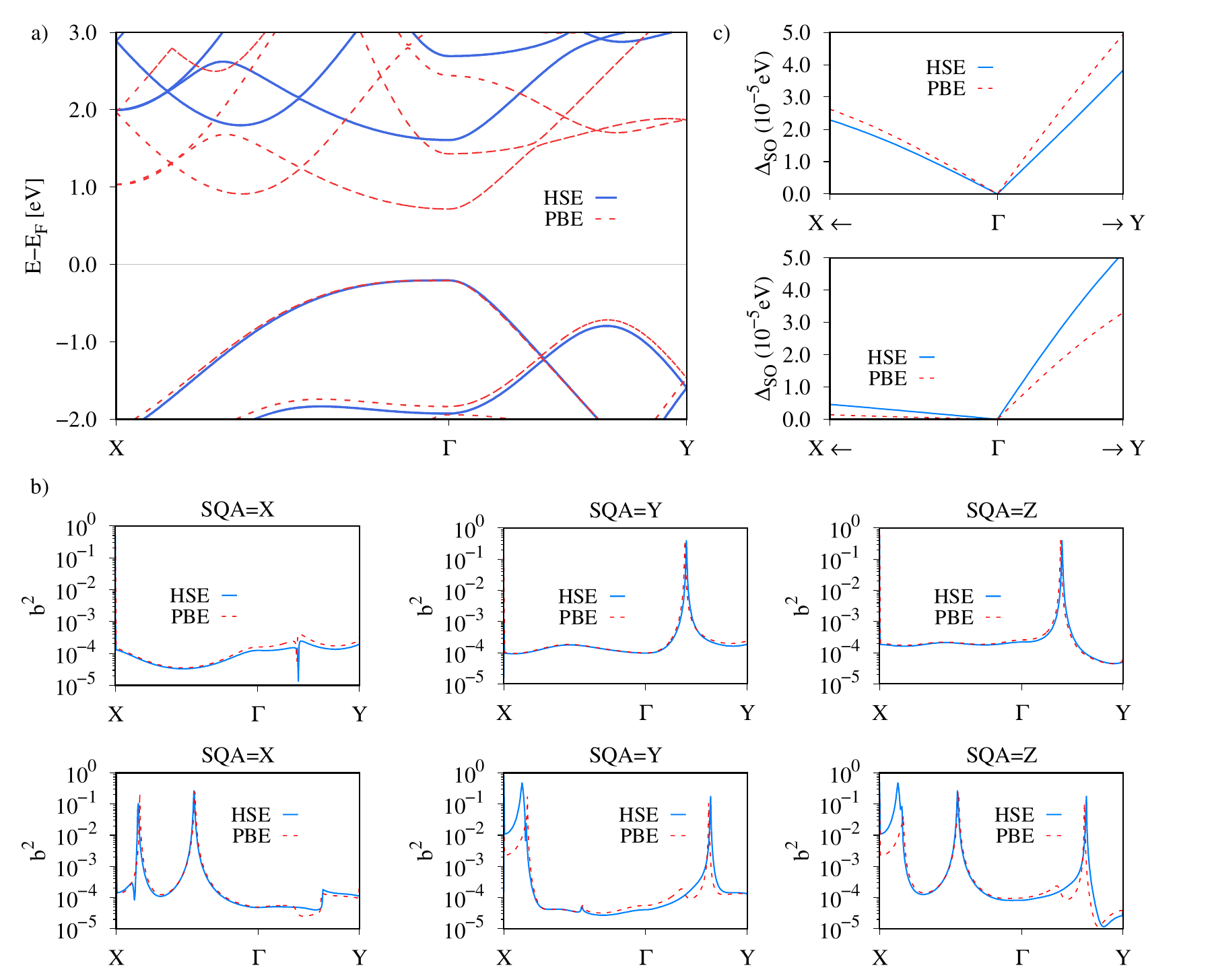}
      \caption{a)~Band structure with HSE\,(blue) and PBE\,(red) functionals for unstrained phosphorene. b)~Comparison spin-mixing parameter $b^2$ along high symmetry points for the next spin quantization axes X, Y and Z in valence and conduction band, respectively. c)~Spin splitting in the vicinity of $\Gamma$ point for valence and conduction band with applied electric field 1V/nm. }
    \label{fig:HSE}
\end{figure*}

In this Section, we investigate the role of the hybrid HSE (\cite{Heyd2003}) exchange-correlation functional and its influence on band structure and spin properties in unstrained phosphorene. The calculations were performed with the Fock exchange contribution 20$\%$ with a q mesh for a sampling of the Fock operator nqx1=nqx2=nqx3=1. The band gap size increases from 0.91\,eV\,(PBE) to 1.81\,eV\,(HSE) without significant changes in the band structure topology, FIG.~\ref{fig:HSE}~a). We observe similar trends in the spin properties of the system, FIG.~\ref{fig:HSE}~b)-c). The anisotropy and strength of the spin-mixing parameter remain the same for HSE and PBE without qualitative changes, while the spin splitting around $\Gamma$ is slightly weaker in the valence band for HSE than for PBE. The situation is the opposite in the conduction band. However, the discrepancies in spin-splitting energies are up to 10$^{-5}$, and the results for HSE and PBE functionals lead to the same findings.

\begin{figure*}[h]
    \centering
    \includegraphics[width=0.9\columnwidth]{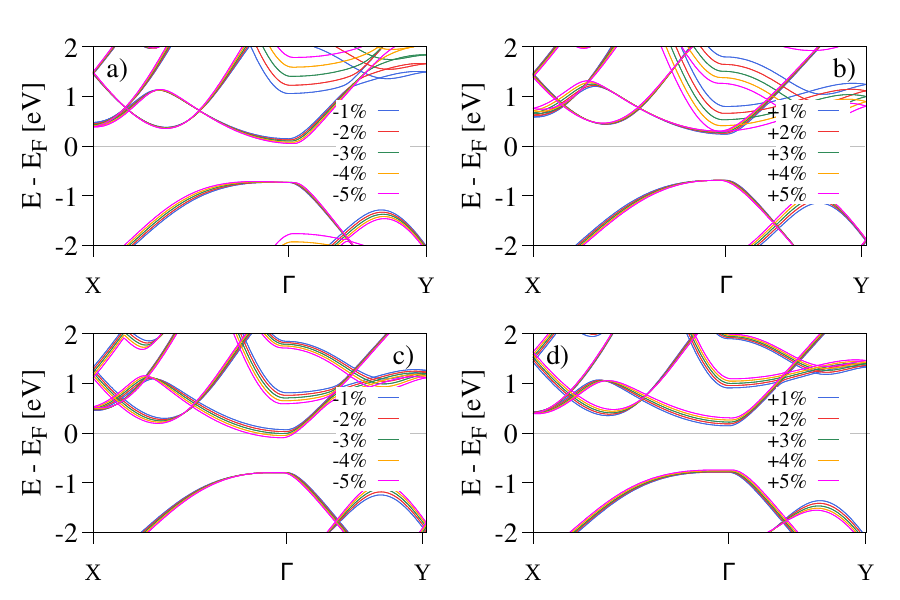}
    \caption{Band structure of strained phosphorene for zigzag a),~b) and armchair c),~d) directions.  }
    \label{figsupp:bnd_strain}
\end{figure*}
\bibliography{bibliography}